\documentclass[sigconf,anonymous=False,natbib=false,nonacm]{acmart}

\usepackage{biblatex}

\RequirePackage[loading]{tracefnt}
\usepackage{algorithmic}
\usepackage{textcomp}
\def\BibTeX{{\rm B\kern-.05em{\sc i\kern-.025em b}\kern-.08em
    T\kern-.1667em\lower.7ex\hbox{E}\kern-.125emX}}
\usepackage{braket}
\usepackage{subcaption}
\usepackage{float}
\usepackage[most]{tcolorbox}
\usepackage[linesnumbered,ruled,vlined]{algorithm2e}

\newcounter{obscounter}
\renewcommand{\theobscounter}{\Roman{obscounter}}
\definecolor{custompurple}{HTML}{53257F}
\newtcolorbox{ObservationBox}[2][]{text width=0.94\linewidth,
colbacktitle=custompurple,enhanced,
attach boxed title to top left={yshift=-2mm,xshift=3mm},
boxed title style={sharp corners},top=6pt,bottom=2pt,
title=#2,colback=custompurple!10!white, left=4pt, right=2pt}
\newcommand{\obs}[1]{
\refstepcounter{obscounter}
\begin{ObservationBox}{\textbf{Observation \theobscounter}}
\par\noindent
#1
\end{ObservationBox}}

\usepackage{tcolorbox}
\usepackage{tabularx}
\usepackage{array}
\usepackage{colortbl}

\newcolumntype{Y}{>{\centering\arraybackslash}X}
\tcbset{tab2/.style={enhanced,fonttitle=\bfseries,fontupper=\normalsize\sffamily,
colback=custompurple!10!white,colframe=custompurple!75!black,colbacktitle=custompurple!30!white,
coltitle=black,center title}}

\tcbset{textmarker/.style={%
        enhanced,
        parbox=false,boxrule=0mm,boxsep=-1mm,arc=0mm,
        outer arc=0mm,left=3mm,right=3mm,top=7pt,bottom=7pt,
        toptitle=1mm,bottomtitle=1mm,oversize}}
\newtcolorbox{noteBox}{textmarker,
    borderline west={3pt}{0pt}{red},
    colback=red!10!white}

\begin{document}

\title{Radiation-Induced Fault Detection \protect\\ in Superconducting Quantum Devices}

\author{Marzio Vallero}
\affiliation{%
  \institution{University of Trento}
  \city{Trento}
  \country{Italy}}
\email{marzio.vallero@unitn.it}

\author{Gioele Casagranda}
\affiliation{%
  \institution{University of Trento}
  \city{Trento}
  \country{Italy}}
\email{gioele.casagranda@unitn.it}

\author{Flavio Vella}
\affiliation{%
  \institution{University of Trento}
  \city{Trento}
  \country{Italy}}
\email{flavio.vella@unitn.it}

\author{Paolo Rech}
\affiliation{%
  \institution{University of Trento}
  \city{Trento}
  \country{Italy}}
\email{paolo.rech@unitn.it}

\renewcommand{\shortauthors}{Vallero et al.}

\begin{abstract}
The quest for universal superconducting quantum computing is hindered by noise and errors.
It has been proven that Quantum Error Correction (QEC) codes will lay at the foundation of fault tolerant quantum computing.
However, cosmic-ray induced correlated errors, which are the most detrimental events that can impact superconducting quantum computers, are yet to be efficiently tackled.
In order to reach fault tolerance, we must also develop radiation aware methods to complement QEC.

In this paper, we propose the first algorithm to effectively exploit syndrome information for the efficient detection of radiation events in superconducting quantum devices at runtime.
We perform a thorough analysis of simulated Rotated Surface codes injecting over 11 million physics-modeled radiation-induced faults.
We consider the properties of the X and Z check bases, the impact of code distance, and the decoder's time to solution constraints.
Our technique detects $100\%$ of injected faults, regardless of the impact's position. Moreover, we accurately identify both the radiation impact centre and the area affected, with an overhead lower than $0.3\%$ the decoding time. Additionally, we use the fault identification information to propose a radiation fault correction technique that improves of up to $20\%$ the output correctness compared to existing decoders.
\end{abstract}




\maketitle

\section{Introduction}
\label{introduction}
Humanity's effort to reach \textit{exascale} compute capabilities was achieved by leveraging GPU accelerators, and there is reason to believe that reaching the next computational threshold will involve hybrid quantum-classical systems.
Despite these promising objectives and almost three decades of research in quantum computing technology, however, current quantum computers are still far from being fault tolerant, menacing the feasibility of this hybrid union.
Among various candidate technologies, superconducting quantum computers are the most promising in terms of manufacturing process maturity and costs, ease of scaleability and gate execution times in the nanosecond regime.
Nonetheless, they are subject to intrinsic noise generated by cross talk amongst proximal qubits, slight differentials in temperature and pressure, and electromagnetic interference.

In order to compensate for intrinsic noise, researchers devised efficient Quantum Error Correction (QEC) codes that, together with classical syndrome decoders, are able to embed and preserve the state of one logical qubit across a set of physical qubits.
Experimental evidence has already proven that it is possible to assemble logical qubits that reach error rates exceeding those of the noisy qubits they are built with.

Unfortunately, as stated by numerous recent papers \cite{Vepsalainen2021, Wilen2021, Acharya2023, Acharya2025,harrington2024synchronous}, the interaction of radiation events with superconducting devices generates chip-wide correlated errors, with event frequency ranging from once per hour to once every ten seconds.
The intensity and persistence of such events is orders of magnitude larger than those of intrinsic noise, effectively erasing the information stored in multiple qubits simultaneously.
In fact, NISQ devices \cite{Preskill_2018} have been already shown to be susceptible to radiation at the algorithmic level \cite{QuFI,Casciola2022cutting,Vallero2024understanding}.
Given that QEC codes are designed to deal with intrinsic noise and not with multiple correlated corruptions, 
there is an argument that current QEC codes are insufficient to address the issue imposed by radiation altogether~\cite{vallero2024efficacy}.
While intrinsic noise suppression in quantum computers is well studied, the areas involving modelling of radiation events, their interaction with QEC codes and the efficiency of classical decoders remain uncharted. Additionally, no effort has yet been done to propose \textit{algorithmic} solutions to detect and eventually correct radiation-induced faults in quantum devices. 

In this paper, we aim to propose innovative software solutions to radiation events in superconducting quantum computers and tackle these three research questions:
\begin{itemize}
    \item \textbf{RQ1}: Can we actively detect high energy events and their area-of-effect in QEC codes at runtime?
    \item \textbf{RQ2}: How do high energy events affect QEC codes and decoders?
    \item \textbf{RQ3}: Can we leverage fault information to improve the decoder's performance?
\end{itemize}

The observations in this article are backed by more than 11 million QEC shots simulated following a physics-derived model of the interaction of radiation with superconducting qubits.
This model considers the distribution of charge deposition from the impinging particle over the quantum chip's surface, and the complete temporal evolution of its transient effects.
The implementation of the fault model is a novel software written as an extension of the STIM library \cite{gidney2021stim}, and is to be disclosed as open source software as part of this article's contributions\cite{repo}.
The choice of a stabiliser-based simulation method is justified by the focus of the paper on QEC codes, as other commonly used quantum circuit simulation methods, such as tensor network or state vector, would require considerably more computational resources at the scales hereby considered \cite{Ahmadzadeh2024,leonteva2025benchmark,Vallero2025state}.

Not only we are able to detect radiation-induced events, but also to identify at runtime their area-of-effect (i.e., the likely set of affected qubits) through a novel QEC-agnostic algorithm that introduces minimal overhead.
Our algorithm, dubbed \textit{Radiation Event Identification} (REI), processes information from the syndrome measurements of a QEC code, correlating it with an internal representation of the qubit's position on the quantum chip.
The incidence of false positives is significantly lower than the incidence of intrinsic noise in the quantum device, which in our experiments is set to $p=10^{-5}$ to replicate the behaviour of near future quantum devices.
We characterise the realtime REI subroutine's predicted area-of-effect and execution time performance over different distance measures of the Rotated Surface code, injecting radiation faults in a set of positions of the quantum chip.
Furthermore, we analyse the embedding of multiple separate quantum error correction codes onto a large quantum chip, benchmarking the efficacy of the Minimum Weight Perfect Matching decoder.
We measure its effectiveness at dealing with syndrome measurements corrupted by radiation, and we observe how the logical error rate of separate logical qubits correlates with their position on the quantum chip.
At last, we measure the performance of a suite of classical graph-based decoders against syndrome measurements corrupted by radiation.
In doing so, we compare such decoders to a novel empirical technique, dubbed \textit{RadMatching}, that aims at correcting radiation's effects just before the decoding step.

This manuscript is structured as follows.
In Section \ref{background_and_related_works} we provide a quick grounding on noise and high energy events in superconducting quantum computers, on QEC and on classical syndrome decoding.
Section \ref{noise_model} describes the formalism and algorithmic implementation of the intrinsic noise model and the radiation-induced fault model.
We detail the QEC codes and decoders object of analysis in Section \ref{exploration_of_design_space}, to later go over the results and answer our research questions in Section \ref{results}.
Conclusions, along details on future investigations, are drawn in Section \ref{conclusions_and_future_works}.

\section{Technical background}
\label{background_and_related_works}
In this section we provide a light background on the topics which are strictly necessary to follow along with the remainder of the article.
It is assumed that the reader is accustomed with the general concepts regarding quantum computing and information.

\subsection{Intrinsic noise and decoherence}
Superconducting qubits encode quantum information in the two-level system of an anharmonic oscillator circuit, which is built using Josephson junctions.
The largest engineering challenge of this technology stems from the requirement of keeping the whole quantum chip well below the critical temperature necessary for observing a supercurrent.
This generally leads to operational temperatures of around $10$ $mK$, but novel technologies show promising performance even at temperature over $200$ $mK$~\cite{hot_qubit}. 
The biggest hurdle is thus to isolate the system enough to preserve its quantum properties, whilst still being able to interact with it.
The ability of a qubit to retain its quantum properties is defined through two measures, the \textit{spin-lattice coherence time ($\tau_1$)} and \textit{spin-relaxation time ($\tau_2$)} \cite{Preskill_2018}.
The former measure, $\tau_1$, is the argument of the inversely decaying exponential function which is used to compute the probability over time of a qubit to have collapsed to the ground state.
Similarly, the $\tau_2$ measure is the argument of another inversely decaying exponential, which instead represents the probability over time for a qubit to have degraded into a classical mixture of states.
Moreover, superconducting quantum computers are subject to imperfections in the application of quantum gates, leading to both state preparation errors and measurement errors.
It is common practice to approximate the ensemble of all these hindering effects through an \textit{intrinsic noise model}, which generally involves appending one random Pauli gate with probability $p$ after each quantum gate or before each measure operator in a quantum circuit. It is important to notice that intrinsic noise is applied on a qubit \textit{independently} of the others. In other words, the error introduced in a qubit is \textit{not} correlated with the error in others.  
The implementation of the intrinsic noise model used in this work is detailed in Section \ref{intrinsic_noise_model}.

\subsection{Radiation-induced events}
\label{high_energy_events}
Following decades of literature on the interaction of cosmic ray derived impinging particles on Silicon, recent experiments \cite{harrington2024synchronous,Wilen2021,Vepsalainen2021} have highlighted that qubits have an even higher radiation sensitivity compared to CMOS devices~\cite{Baumann2005}.
In a quantum chip, an impinging particle deposits energy in the Silicon substrate along a trail and leaves behind a large amount of electron-hole pairs~\cite{Cardani2023}.
This transient potential is always sufficient to break the feeble bond, on the order of a few $meV$, that holds together Cooper pairs in the superconducting elements of the quantum chip.
In turn, this gives rise to secondary products such as phonons, which are responsible for the long lasting effects of radiation faults~\cite{Martinis2021}.
However, whilst classical CMOS are only affected by radiation when the deposited energy is larger than the critical charge threshold of the transistor and remain unaffected otherwise~\cite{Baumann2005}, the state of qubits can be disturbed even by the breaking of a single Cooper pair~\cite{casagranda2025tns,Cardani2023}.

In the event of a fault, the stored quantum information erasure is triggered, fundamentally reducing the coherence time $\tau_1$ of the qubit~\cite{Vepsalainen2021}.
This makes quantum devices prone to errors generated not only by relatively rare highly interacting particles such as neutrons ($13 \ n/cm^2/h$ at sea level), but also from the overly abundant flux of muons ($60 \ muons/cm^2/h$ at sea level) \cite{Jedec2006,casagranda2025tns}. 
Experimental evidence has shown that radiation faults have been registered with frequencies ranging from once every tens of seconds to once per hour, depending on a multitude of factors, such as environmental shielding and gap engineering of the Josephson junctions \cite{Acharya2023,Acharya2025,Wilen2021}.

Critically, since the deposited charge quickly diffuses in the substrate, the documented area-of-effect of these events usually involves most, if not all, of the qubits present on the quantum chip.
This goes in stark contrast with the error rates of classical terrestrial computing systems, where a detrimental radiation event usually affects just a few bits of memory over an entire supercomputer with trillions of transistors.
The nature of an impinging particle's byproducts makes them quite persistent, with transients lasting from a few milliseconds to upwards of tens of seconds\cite{mcewen2022resolving}.
When considering that a round of a QEC code lasts from a few hundred to a few thousand \textit{nanoseconds}\cite{Acharya2023}, depending on code distance, this means that a large portion of correction cycles will be affected by radiation.
In simpler terms, an impinging particle induces spatially correlated errors, with higher intensity at the impact locus and overall decreasing magnitude over time.
Further details on the implementation of our physics-derived radiation fault model are provided in Section \ref{radiation_fault_model}.

Current solutions to tackle radiation range from building isolated or "suspended" circuits\cite{Junger2025suspendedqubits}, in order to avoid contact between the qubit and the Silicon substrate or other qubits, to specialised classical hardware routines aimed at correcting small uncorrelated errors.
Furthermore, to improve the device's environmental shielding and reduce the mean incoming particle flux, quantum computers are being buried hundreds of meters underground\cite{Cardani2021,Loer2024underground}.
Albeit all of these approaches have been proven to provide some benefits, most of them are hardly scaleable and expensive.
There is an urgent need to find software level solutions to this issue, as this can not be solved purely from a hardware standpoint.
The first step to do so is to sharply distinguish when and where radiation affects QEC codes.

\subsection{Quantum Error Correction}
Classical error correction codes, in a simplified picture, make use of data replication and parity measurements to reduce the error rate associated to memory devices or communication protocols~\cite{Chatterjee2023qec_for_dummies}.
However, the picture for quantum information is more complex, since one can not copy an arbitrary quantum state, nor read its information without destroying it \cite{Wootters1982nocloning}.
These limitations have been tackled through the usage of stabiliser measurements and data replication, which make up the fundamental building blocks of stabiliser-based QEC codes~\cite{gottesman1997qec,Chatterjee2023qec_for_dummies}.
As such, a simplified QEC code makes use of a set of \textit{data} qubits, which typically embed the replicated information of one logical qubit, and a set of \textit{stabiliser} qubits, which are responsible for computing the joint parity of a subset of neighbouring data qubits through projective measurements on a given basis~\cite{gottesman1997qec,Chatterjee2023qec_for_dummies}.
Each stabiliser shares one data qubit with \textit{at least} one other stabiliser, and computes the joint parity of \textit{at least} two data qubits.
The set of stabilisers can be used as the generators of the parity check matrix of the QEC code.
The relations in the parity check matrix can also be represented in a Tanner graph, a connected graph where each vertex represents one stabiliser qubit and each edge represents an error mechanism between two stabilisers.
Edges may thus be due to error mechanisms on a data qubit shared between two stabilisers qubits, or due to error mechanisms on the stabiliser itself.
The Tanner graph in STIM is represented as a \texttt{DetectorErrorModel} (DEM) object \cite{gidney2021stim}.
Our idea, further detailed in Section \ref{decoder_performance_comparison}, is to extract information on the radiation event to be later exploited before the decoding step.

Quantum error correction codes are generally parameterisable by \textit{code distance} $d$ and \textit{number of repetitions} $r$.
The distance measure is directly proportional to the number of data qubits over which quantum information is replicated.
The number of repetitions identifies how many times the stabilisers in the QEC code must be re-measured over time to complete a round of correction, and generally scales as $O(d)$.
As such, a single stabiliser qubit may be measured multiple times during the execution of a single round of correction, producing separate stabiliser measurements.

Once a QEC measurement round has been completed, the parity information of all the stabilisers is collected into a \textit{syndrome measurement}, a classical vector of boolean variables.
A \textit{true} value represents an odd parity for that stabiliser, whilst a \textit{false} value represents an even parity.
In this context, stabilisers that have measured an odd parity are labelled as \textit{defects} of the syndrome.

For the purposes of this article, we consider the Rotated Surface code~\cite{Bombin2007rotatedsurfacecode,Kovalev2012rotatedsurfacecode}, also known as checkerboard code for the characteristic pattern of its stabiliser generators, which is an alternative version of the surface code which has been rotated by $\pi/4$.
This makes it so that data qubits are placed on the vertices of a square lattice, and the X and Z stabiliser qubits are placed at the centre of the plaquettes in a checkerboard pattern.
The Rotated Surface code of distance $d$ maps the logical state of a single qubit onto $2d^2-1$ qubits, with $d^2$ data qubits and $d^2-1$ stabiliser qubits.

\subsection{Syndrome decoding}
\label{other_decoders}
The syndrome measurement obtained from a round of QEC must be processed in order to identify which error mechanisms in the Tanner graph might have the highest chance to be responsible for the syndrome's defects.
This can be visualised as finding the set of edges connecting defect pairs in the Tanner graph which covers the data qubits that are most likely to have triggered a defect, finding a \textit{matching}.
However, not all syndrome measurements have even parity, as such one odd stabiliser might be left unmatched.
As such, the Tanner graph is expanded with a boundary hyper-vertex, that acts as an escape route that can match any number of defects.
The process of finding the most probable set of defect-joining edges, known as maximum likelihood decoding, is known to be NP-hard, and thus alternative and faster workaround techniques have been developed by the community.
The idea behind these alternative approaches is that intrinsic noise leads to uncorrelated pairwise defects, as such the shortest path between two unmatched syndromes should cover the most probable defect source.
This information is then used to both read out the logical state of the QEC code and to correct the sources of error.

Many decoding techniques are available in the literature, ranging from graph matching algorithms~\cite{pymatching}, to Machine Learning implementations that specialise on specific noise profiles and QEC codes~\cite{Acharya2025,Bausch2024ml_qec_decoder,Sweke2021ml_qec_decoder}, to Tensor Network contraction approaches~\cite{Acharya2025,Ferris2014tn_qec_decoder,Chubb2021tn_qec_decoder}.
Generally, graph-based algorithms boast the best compromise between fast time to solution and excellent accuracy~\cite{Battistel2023decoder_review,Acharya2025}.
In fact, the decoding step is subject to very strict time constraints, since it needs to keep pace with the syndrome measurement generation rate of real quantum computers to avoid the buildup of an exponential backlog of syndromes to be processed.
To fulfill these requirements, Tanner graph based algorithms have been selected as prime candidates for the construction of \textit{ad hoc} ASIC systems~\cite{Fowler2015asic,Battistel2023decoder_review,Barber2025asic_decoder}.
Given the scale of our work, we will only evaluate the performance of Tanner graph based approaches, which include the following.

\noindent
\textit{Minimum Weight Perfect Matching}.
This decoder uses a variation of Edmond's Blossom algorithm \cite{Edmonds1965blossom} to find a matching of the Tanner graph which is minimal, according to the sum of the weights in the matching edge set, where the weight of one edge is the distance between its vertices.

\noindent
\textit{Belief Matching}.
The Belief Matching (BM) decoder is composed of two subroutines, with a Belief Propagation (BP) step immediately followed by a MWPM step \cite{higgott2023improveddecodingcircuitnoise}.
Similarly to the Belief Find decoder, the syndrome induced Tanner graph is first tentatively decoded by the BP subroutine.
In case of failure, the information from the BP step is used to set the weights for the MWPM step.

\noindent
\textit{Union Find}.
This decoder uses the Union Find (UF) algorithm applied to the parity check matrix of the QEC code to infer the estimated correction vector from the syndrome data \cite{Delfosse_2021, delfosse2021unionfinddecoderquantumldpc}.

\noindent
\textit{Belief Find}.
This decoder is composed of two routines.
At first, the Belief Propagation (BP) algorithm is used to try to find a correction vector.
If BP does not converge, the edge weights obtained from that step are fed to the Union Find algorithm \cite{higgott2023improveddecodingcircuitnoise}.

The implementation of the BP and UF subroutines is provided by the \textit{ldpc} library \cite{Roffe_LDPC_Python_tools_2022}, while the MPWM subroutine is provided by the \textit{PyMatching} library \cite{pymatching}.

\section{Quantum chip, Noise and Fault Modelling}
\label{noise_model}
This section details the general characteristics of our superconducting quantum chip model, the algorithmic implementation of the intrinsic noise model for a general superconducting quantum computer and the radiation fault model used in all of the analyses of Section \ref{results}.

\begin{figure}
    \centering
    \includegraphics[width=.8\linewidth]{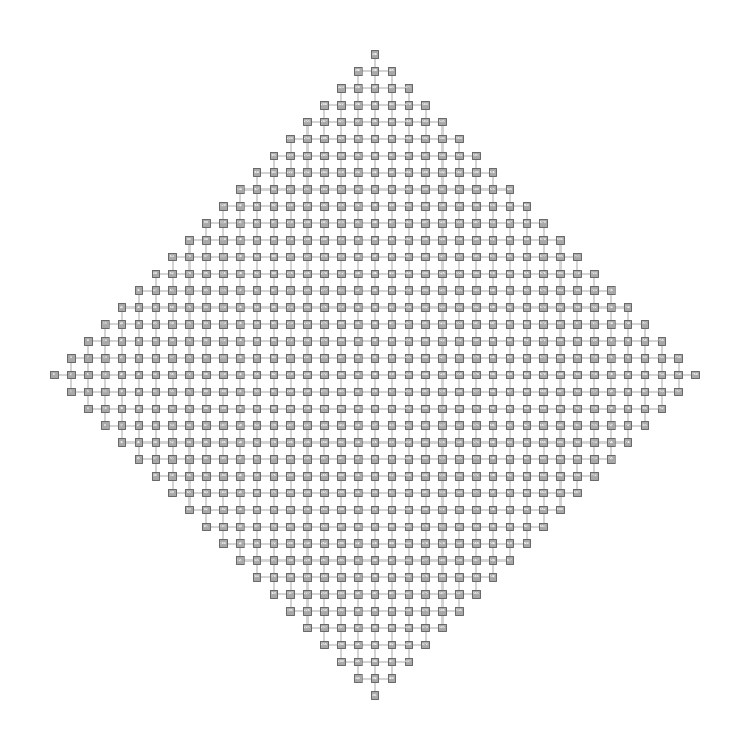}
        \caption{\textbf{Quantum chip topology}. The structure of the quantum chip we modelled, generated from the repetition of a cross-shaped minimal connected component over 20 diagonal rows and 20 diagonal columns, with a total of 760 qubits.}
    \label{chip_topology}
\end{figure}

\subsection{Quantum chip model}
\label{device_model}
We modelled a generic quantum chip, which is able to execute any quantum gate amongst the ones available in the STIM library.
This quantum chip model follows the reported topology and hardware properties of top of the line quantum computers at the time of writing \cite{Acharya2025,Bravyi2024GrossCode}.
We defined the characteristic $\tau_1$ coherence time of this simulated quantum chip to be of $85 \mu s$, and we set the gate durations to be of $25 ns$ for single qubit gates, $32 ns$ for two qubit gates and $58 ns$ for measure and reset gates.
Moreover, the topology of the interconnections between qubits has been selected so that the QEC codes that we took into consideration could be easily mapped on the quantum chip without incurring in the need of SWAP operations.
The mapping is found through an homomorphic subgraph solver, by searching the interconnection structure of a given quantum circuit into the quantum chip's topology.
The topology is built from the repetition over two axes of a single connected component by a number or \textit{rows} and \textit{columns}.
An example of the topology of the quantum chip, as seen from a top-down view, is provided in Figure~\ref{chip_topology}.

Although these characteristics are representative of current quantum devices, thanks to the modular and open source design of the code library that we developed \cite{repo}, any current and future quantum chip can be easily modelled and tested as well.

\begin{table}[!t]
\centering
\renewcommand{\arraystretch}{0}
\begin{tcolorbox}[tab2,tabularx={c||Y|Y|Y|Y|Y|Y},title=Instrinsic noise model,boxrule=0.6pt]
\small Set \rule[1ex]{0pt}{1.2ex} & $I$ & $\mathcal{P}_e$ & $II$ & $I\mathcal{P}_e$ & $\mathcal{P}_eI$ & $\mathcal{P}_e\mathcal{P}_e$\\[.2ex]\hline\hline
$\mathcal{Q}_1$ \rule[1ex]{0pt}{1.2ex} & $1-\frac{p}{10}$ & $\frac{p}{30}$ & $-$ & $-$ & $-$ & $-$ \\[.2ex]\hline
$\mathcal{Q}_2$ \rule[1ex]{0pt}{1.2ex} & $-$ & $-$ & $1-p$ & $\frac{p}{5}$ & $\frac{p}{5}$ & $\frac{3p}{5}$ \\[.2ex]\hline
$\mathcal{M}$ \rule[1ex]{0pt}{1.2ex} & $1-5p$ & $\frac{5p}{3}$ & $-$ & $-$ & $-$ & $-$  \\[.2ex]\hline
$\mathcal{R}$ \rule[1ex]{0pt}{1.2ex} & $1-2p$ & $\frac{2p}{3}$ & $-$ & $-$ & $-$ & $-$   \\[.2ex]
\end{tcolorbox}
\caption{The probability for each Pauli error to be triggered for set of quantum gates, in relation to the noise error rate $p$, which reflects the average error rate of a real superconducting quantum computer, adapted from the SI1000 noise model \cite{Gidney2022intrinsicnoisemodel}.}
\label{instrinsic_noise_model_table}
\end{table}

\begin{figure*}[!t]
    \centering
    \includegraphics[width=0.8\linewidth]{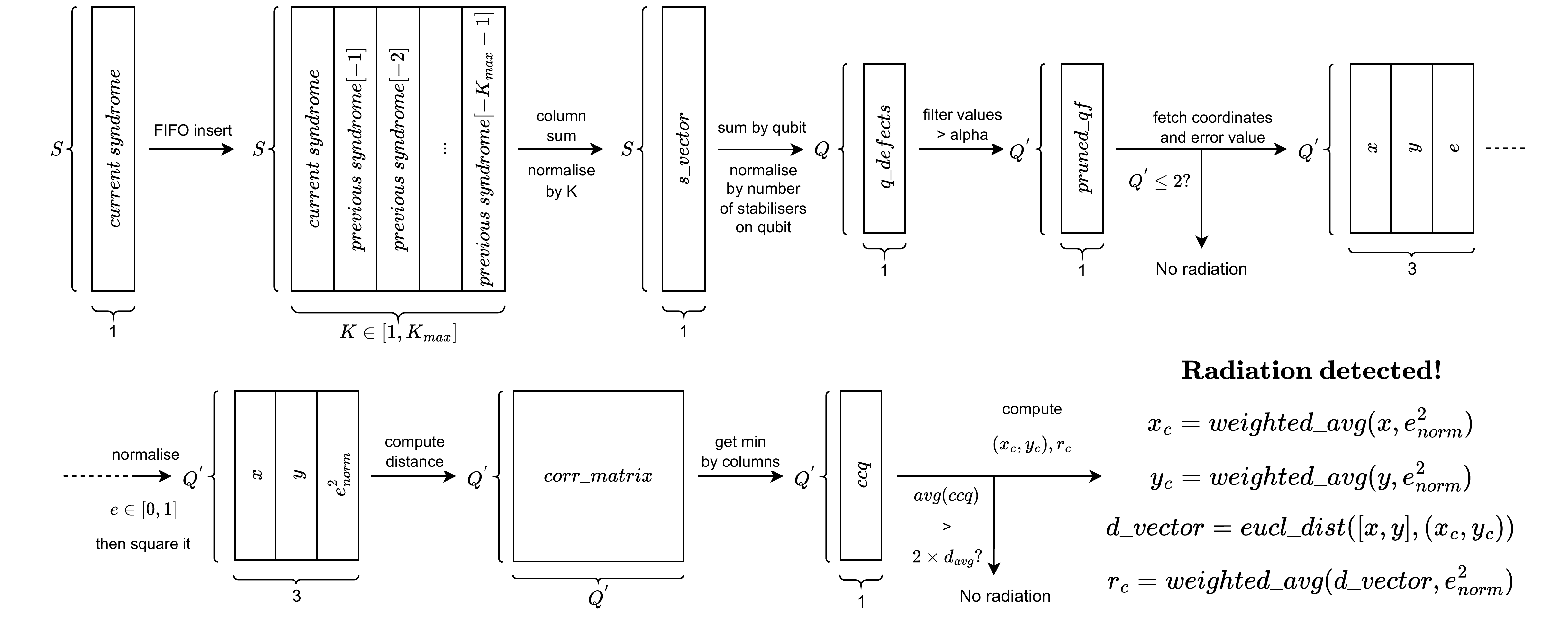}
        \caption{The information processing scheme of the REI subroutine.}
    \label{rei_algorithm_scheme}
    
\end{figure*}

\subsection{Intrinsic noise model}
\label{intrinsic_noise_model}
Superconducting quantum computers are subject to an ensemble of physical phenomena that hinder the accuracy of quantum gate operations.
Given the complexity of all the variables at play, QEC codes are usually tested against artificial noise that optimally approximates a real quantum computer's behaviour.
It is common practice in the literature to compose such \textit{intrinsic noise models} through the usage of \textit{Pauli operators}, under the umbrella terminology of \textit{depolarisation error models} \cite{Georgopoulos2021}.
They are parameterised by a \textit{noise error rate} $p$, which is meant to match the average measured error rate of a real quantum computer.
The intrinsic noise model we considered is a variation of the \textit{superconducting-inspired 1000 ns cycle} (SI1000) noise model \cite{Gidney2022intrinsicnoisemodel}.
We identify four sets of quantum operations: the set of single qubit gates $\mathcal{Q}_1$, the set of two qubit gates $\mathcal{Q}_2$, the set of reset gates $\mathcal{R}$ and the set of measurement operators $\mathcal{M}$.
Noise is thus introduced in the quantum circuit operator that probabilistically triggers either an $I$ operator or an operator in the set $\mathcal{P}_e \doteq \{X,Y,Z\}$ after each qubit involved in a quantum gate or measurement operator, governed by the model's \textit{noise error rate}.
In the case of the $\mathcal{Q}_2$ set, the noise operator is the tensor product of two Pauli operators, which can either be $I$ or an operator in the $\mathcal{P}_e$ set.
Each of those noise operators is then triggered independently, giving rise to uncorrelated errors across qubits and over time, following the relations in Table \ref{instrinsic_noise_model_table}.

Considering that the coherence time of modern superconducting quantum computers is orders of magnitude longer than the execution time of a single QEC code round, it is reasonable to assume that the noise error rate $p$ remains constant throughout the execution of the simulations.
In all of the simulation results discussed in Section \ref{results}, we consider an intrinsic noise rate $p=10^{-5}$.

\subsection{Radiation fault model}
\label{radiation_fault_model}
Following the experimental characterisations provided in Section \ref{high_energy_events}, we know that high energy events disturb the quantum information stored in multiple qubits, ultimately reducing their coherence time.
This behaviour has been modelled via a probabilistic \textit{Y gate} prepended to each quantum gate operation, that induces a loss of coherence by erasing the information in the qubit according to a probability $p_{q_i}$.
This probability depends on three factors, namely the time elapsed between the previous gate operation on a qubit and the current gate operation on that same qubit, the spatial distance between the impact locus of the high energy event and the qubit, and the time elapsed since the beginning of the high energy event.
From this, it follows that the high energy event's intensity depends on the quantum circuit's structure, the architecture of the quantum chip and time.

Over the \textbf{temporal dimension}, the probability for a qubit to undergo decoherence depends on the $\tau_1$ measure and the time elapsed between the previous and the current quantum gate $\Delta t_g$.
High energy events significantly reduce $\tau_1$, spiking in intensity as soon as energy is deposited and gradually wearing off over time.
This reduction has been modelled through the $\tau_{rad}(t)$ function, detailed in Equation \ref{tau_rad}, that is parameterised over the time $t$ at which the quantum gate is applied, and depends over three constants: the characteristic $\tau_1$ of the quantum computer, the wall-clock time at the start of the high energy event $t_{rad}$ and the total duration of the high energy event $\Delta t_{rad}$.
\begin{equation}
\label{tau_rad}
\tau_{rad}(t) = \tau_1 e^{10 \left( \frac{t - t_{rad}}{\Delta t_{rad}} - 1 \right)}
\end{equation}
At the locus of the high energy event, the probability $p_{root}$ of the fault to trigger is computed through the $T(\Delta t_g, t)$ function, detailed in Equation \ref{time_evolution}, which also depends on the current value of $\tau_{rad}(t)$.
\begin{equation}
\label{time_evolution}
T(\Delta t_g,t) = 1 - e^{-\frac{\Delta t_g}{\tau_{rad}(t)}}
\end{equation}
Following the device model described in Section \ref{device_model}, the time required to execute a round of quantum error correction will range from a few hundred to a few thousand nanoseconds, depending on the depth of the quantum circuit that encodes the QEC code.

Over the \textbf{spatial dimension}, the charge deposited by a high energy event spreads across the quantum chip, where the intensity of the radiation fault becomes lower the further away a qubit is from the impact point.
Knowing the planar coordinates of each qubit on the quantum chip's architecture, one can compute the Euclidean distance over two dimensions $\Delta s$ from the locus of impact to any qubit on the quantum chip.
To model the decreasing intensity of the radiation fault over the quantum chip, we use the $S(\Delta s)$ function, which decreases as the inverse square of the distance from the locus of impact, as shown in Equation \ref{space_evolution}.
\begin{equation}
\label{space_evolution}
S(\Delta s) = \frac{1}{(\Delta s + 1)^2}
\end{equation}

Piecing Equations \ref{tau_rad}, \ref{time_evolution} and \ref{space_evolution} together lets us define the $P(\Delta s, \Delta t, t)$ function, which represents the probability $p_{q_i}$ of the qubit of index $i$ to undergo a radiation-induced erasure error, as detailed in Equation \ref{fault_probability}.
\begin{equation}
\label{fault_probability}
P(\Delta s, \Delta t_g, t) = S(\Delta s) T(\Delta t_g,t)
\end{equation}

\obs{
The radiation-induced fault intensity depends on a qubit's distance from the source of radiation, the time since the beginning of the radiation event and the idle time between gates.
}

\begin{algorithm}[!t]
\small
\SetAlgoLined
    \KwIn{\texttt{syndrome}}
\KwOut{\texttt{(x, y), radius}}
\texttt{\noindent
\uIf{$s\_matrix$.rows = $0$}{
    $s\_matrix$ $\gets$ $syndrome$\\
}
\Else{
\If{$s\_matrix$.rows $\ge$ $K_{max}$}{
    $s\_matrix$.delete(row=$0$)\\
}
$s\_matrix$.append($syndrome$)\\
}
$s\_vector$ $\gets$ $s\_matrix$.sum(axis="cols") \\
$s\_vector$ $\gets$ $s\_vector / s\_matrix$.rows \\
$q\_defects$ $\gets$ vector(shape=$(Q,)$, init\_value=$0$)\\
\For{($q$, $s$) $\in$ $qubit\_to\_stabilisers$}{
    $q\_defects$[$q$] $\gets$ $s\_vector$[$s$].sum()$/s$.rows\\
}
$alpha$ $\gets$ $1 / ((rounds + 1) s\_matrix$.rows)\\
$pruned\_qf$ $\gets$ $q\_defects$[$q\_defects > alpha$]\\
\If{$pruned\_qf$.rows $\leq 2$}{
    \Return NULL\\
}
$xye$ $\gets$ matrix(rows=$pruned\_qf$.size, cols=$3$)\\
\For{$(i, (q, e)) \in pruned\_qf$}{
    $xye$[$i, :$] $\gets$ \texttt{vector}([$q.x, q.y, e$])\\
}
$range$ $\gets$ $xye$[$:, 2$].max$ - xye$[$:, 2$].min\\
\If{$range$ $\ne$ $0$}{
    $xye$[$:, 2$] $\gets$ ($xye$[$:, 2$]$ - xye$[$:, 2$].min)$/range$ \\
}
$corr\_matrix$ $\gets$ \texttt{euclidean\_dist}($xye[:,0:2]$, $xye[:,0:2]$, ignore\_self=\texttt{True})\\
$closest\_corr\_qubits$ $\gets$ $corr\_matrix$.min(axis="cols")\\
$correlation\_factor$ $\gets$ \texttt{avg}($closest\_corr\_qubits$)\\
\If{$correlation\_factor$ $>$ $2*device\_avg\_min\_dist$}{
    \Return NULL\\
}
$xye$[$:, 2$] $\gets$ power($xye$[$:, 2$], 2)\\
$x$ $\gets$ avg($xye$[$:, 0$], weights=$xye$[$:, 2$])\\
$y$ $\gets$ avg($xye$[$:, 1$], weights=$xye$[$:, 2$])\\
$d\_vector$ $\gets$ euclidean\_dist($xye$[$:, 0:2$], \texttt{vector}([$x,y$]))\\
$radius$ $\gets$ $2 *$ \texttt{avg}($d\_vector$, weights=$xye$[$:, 2$])\\
\Return (($x$, $y$), $radius$)
}
\caption{Radiation Event Identification (REI)}\label{rad_detection}
\end{algorithm}

\section{Radiation Event Identification (REI)}
\label{exploration_of_design_space}
The main idea behind the Radiation Event Identification (REI) subroutine is to detect and characterise a radiation-induced fault making use of information regarding the device's architecture, the QEC code's mapping on the device, and a dynamic backlog made of the previous syndrome measurements.
This algorithm has been explicitly designed to be agnostic of the QEC class, making it a widely applicable subroutine for quantum error correction decoders.
In Section~\ref{decoder_performance_comparison} we also propose how to use this information to move towards radiation-aware \textit{decoding}.
With \textit{identification} we mean both the ability of the algorithm to discern whether a radiation event is happening in real time, and that of knowing the approximate centre of the impact location and its extension over the quantum chip. 
REI is comprised of three steps: (1) radiation-fault detection, (2) radiation impact location identification, and (3) area affected by radiation. The latter two steps are necessary to evaluate if the unaffected qubits can still be used to reconstruct the information regardless of the radiation-induced corruption.

Given the radiation fault model presented in Section \ref{radiation_fault_model}, we know that high energy events show correlations in space, over neighbouring qubits on the quantum chip, and in time, across time spans that last for thousands of QEC shots.
The approach is presented in Algorithm \ref{rad_detection}, jointly with the information processing scheme of Figure \ref{rei_algorithm_scheme}.

Since syndrome measurements are processed sequentially over time, the decoder keeps track of the last $K_{max}$ syndromes with a FIFO policy, in order to correlate radiation errors in time.
The backlog of $K\in[1,K_{max}]$ measured syndromes of size $S$ is stored in a dynamic matrix \textit{s\_matrix} of size $K \times S$.
Whenever a new syndrome measurement is provided, the oldest syndrome measurement is discarded, and the current syndrome measurement is appended as the last row of \textit{s\_matrix}.
This latter matrix is then reduced to a vector \textit{s\_vector} of size $S$ by summing over all of the column values, and then normalised by the number of rows in \textit{s\_matrix}.
We instantiate a vector \textit{q\_defects} of size $Q$, which is the number of physical qubits in the QEC code considered.
Each stabiliser measurement in a QEC code is hosted on one specific physical qubit.
The \textit{qubit\_to\_stabilisers} dictionary maps each physical qubit index $q$ to a vector of stabiliser indexes $s$ that are hosted on $q$.
Iterating over all physical qubits $q$ used by the QEC code, we store in \textit{q\_defects}[$q$] the sum of the elements in \textit{s\_vector} with indexes $i \in s$, normalised by the cardinality of $s$.
The \textit{q\_defects} vector is pruned, keeping only those elements which are strictly larger than $alpha=1/((rounds+1) S)$, and its results are stored in the dictionary $pruned\_qf$.
If the number of elements in $pruned\_qf$ is smaller than two, no centroid can be found, and the subroutine returns.
Otherwise, we instantiate a matrix $xye$ with rows equal to the number of elements in $pruned\_qf$, and three columns.
Iterating with index $i\in[0,pruned\_qf.size]$ over the $(q,e)$ pairs in $pruned\_qf$, we set the i-th row of $xye$ as the $x$ and $y$ coordinate of qubit $q$ and its defect incidence rate $e$.
We then compute the Euclidean distances amongst all qubit planar coordinates in $xye$, storing them in a distance matrix $corr\_matrix$.
In doing so, we ignore the diagonal elements, which represent the distance of each qubit with itself.
We then filter the minimum value for each column in the $corr\_matrix$ and collect them in the $closest\_corr\_qubits$ vector, then we average such vector to compute the $correlation\_factor$.
If the latter is found to be strictly greater than the average minimum euclidean distance between all qubits in the quantum chip's topology, then the events are considered uncorrelated and the subroutine returns without detecting a radiation event.
\begin{figure*}[!ht]
    \centering
    \includegraphics[width=.85\linewidth]{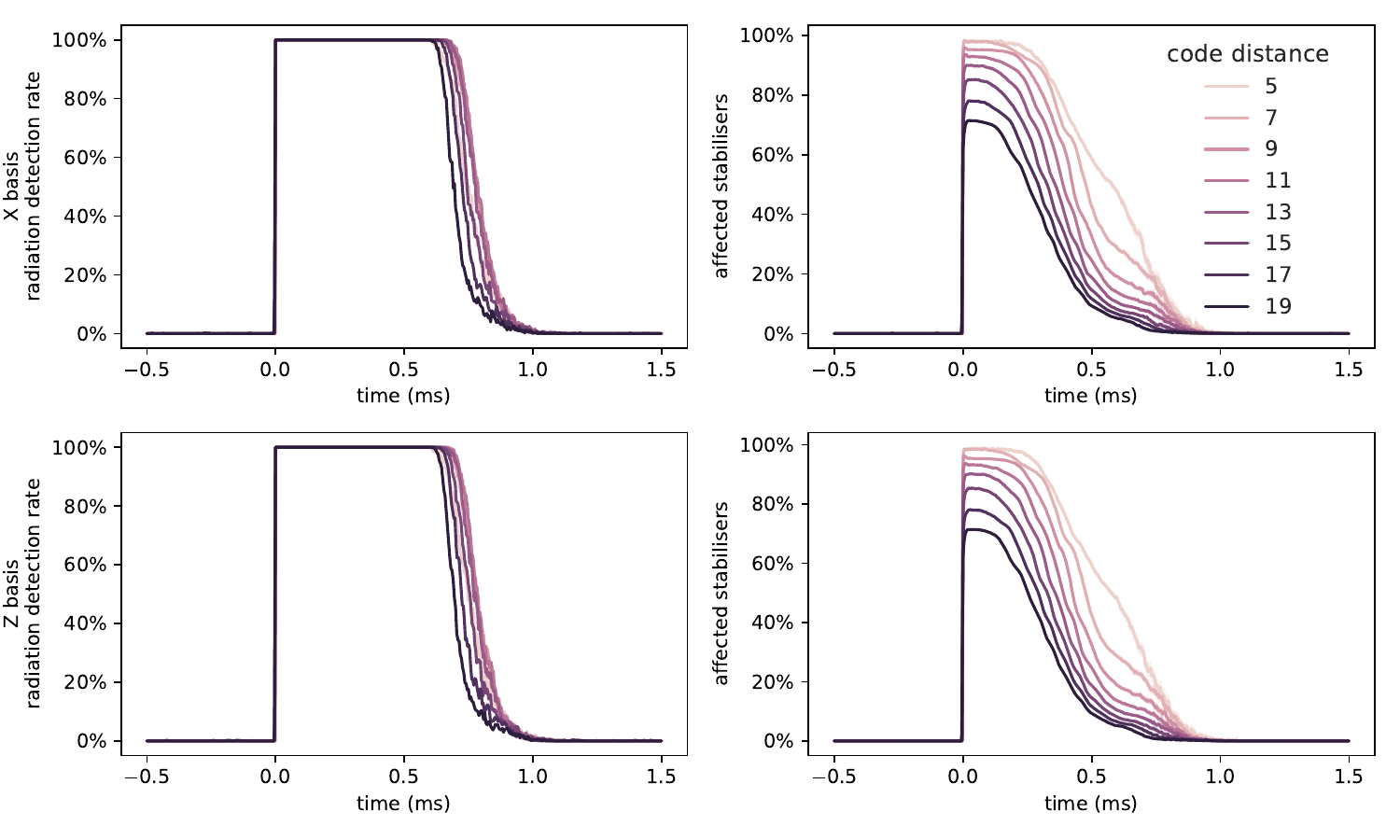}
        \caption{\textbf{Code distance relation with radiation area-of-effect detection}. We consider the Rotated Surface code, both in the X-basis and the Z-basis correction steps, at code distances varying from 5 to 19. Statistics are drawn over 512 samples.}
    \label{code_distance_detection}
    
\end{figure*}
\noindent
Otherwise, the physical $x,y$ coordinates of each physical qubit, the coordinates of the locus of the high energy event are computed as the average over the $x$ and the $y$ coordinates of all qubits, weighted by the error column $xye$[$:,2$] which had been previously elevated to the power of two.
The radius of the high energy event is computed as twice the average weighted euclidean distance between each qubit and the newfound coordinates of the radiation error locus.

\obs{
We can identify a circular threshold that encloses radiation's area-of-effect, which shrinks over time.
All equidistant qubits experience on average the same radiation-induced fault intensity.
}

\section{Results}
\label{results}
To show the efficacy and efficiency of our proposed subroutine and its applications, we devised a set of five separate analyses, considering variable code distances for the Rotated Surface code, different loci for radiation faults, multiple Tanner graph based decoder implementations, and configurations of the mapped surface codes on the quantum chip's architecture.

\begin{figure*}[!ht]
    \centering
    \includegraphics[width=.9\linewidth]{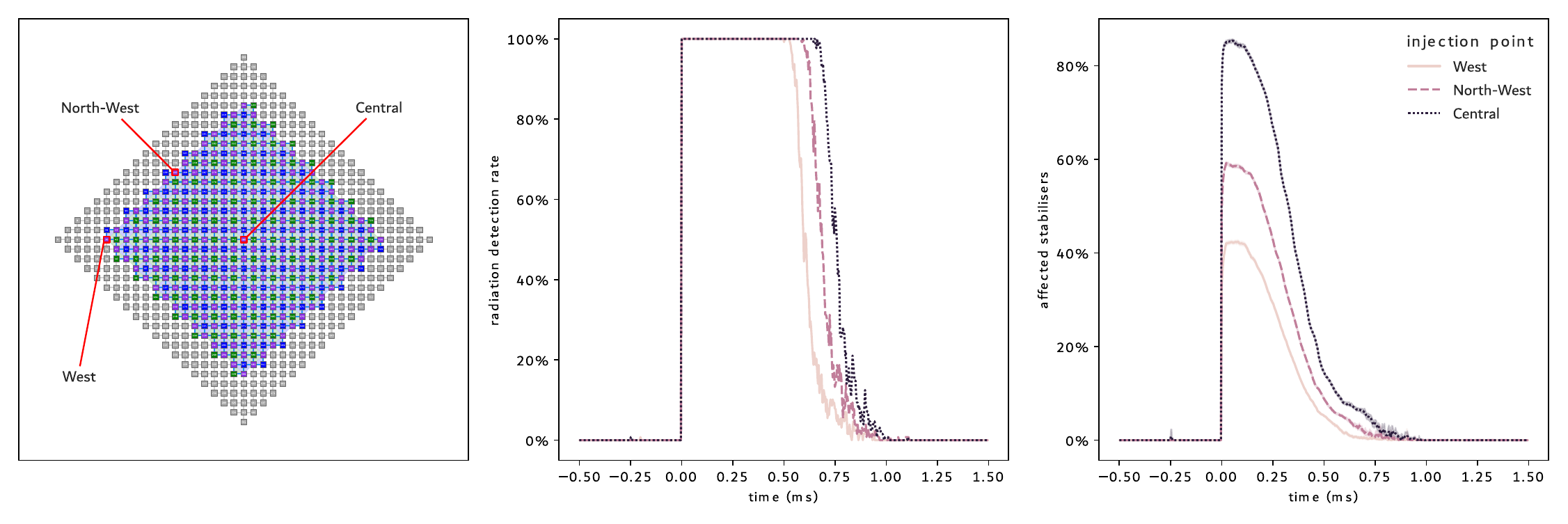}
        \caption{\textbf{Radiation area-of-effect detection}. We consider the Rotated Surface code (Z-basis) of distance 15 with 15 repetitions per round of correction. We inject in three different locations (left, outlined in red), over separate simulations. Then we plot the detection rate of radiation events (centre), and the estimated ratio of stabilisers affected over time by the radiation event (right). Statistics are drawn over 128 samples.}
    \label{rad_aoe}
    
\end{figure*}

\subsection{Code distance impact on detected area-of-effect}
The subject of this analysis is to compare the REI subroutine efficacy on the Rotated Surface code against the same simulated radiation event at different code distances.
To keep the analysis as interesting as possible, we selected as the radiation locus the centre of the quantum chip, which is the worst case scenario for such an event.
We consider a radiation event beginning at time zero, with an overall duration of $1$ $ms$, meaning that the intensity of the fault after that time is strictly lower than the intrinsic noise rate of the quantum device.
We draw statistics over 128 independent syndrome sequences over time.

In Figure \ref{code_distance_detection}, we compare the REI's outputs on both the X-basis and the Z-basis correction passes of the Rotated Surface code, on the first and the second row respectively.
In the first column, we notice how the radiation detection rate, that is the incidence by which the REI subroutine successfully spots a radiation event, suffers from no false positives outside of the time window of the fault.
Moreover, the identification of the fault is triggered sharply as soon as it begins, regardless of the code distance and the basis considered.
Notably, the tail of the radiation fault, which is exponentially less intense at the end than at the beginning of the event, stops being detected slightly earlier in higher distance codes.

\obs{
Radiation event identification is code and error-basis agnostic, and can be identified at runtime and separated from regular intrinsic noise.
}

In the two rightmost plots of Figure \ref{code_distance_detection}, we consider the ratio of stabilisers affected by the radiation event over the total number of stabilisers in the surface code.
Given the properties of radiation faults described in Section~\ref{radiation_fault_model}, we expect the affected area to be circle-shaped and centred on the injection point.
In all of the considered experiments, all code configurations are subject to a simulated radiation fault of equal position and intensity, thus inducing comparable error rates in the same qubits across simulations.
Notably, the ratio of corrupted stabilisers is inversely proportional to the code distance, as they are more closely compacted around the fault's source.
In fact, codes which use a lower number of qubits will inevitably incur a higher ratio of failing stabilisers, as the area of effect of radiation is the result of a physical phenomenon, unaffected by the choice of QEC code.
Inversely, using a larger number of resources will in turn reduce the percentage of affected stabilisers, as the QEC code will have more stabilisers "to spare".
Once again, the choice of the Rotated Surface code's basis is transparent to the measured metric.

\obs{
Larger distance Rotated Surface codes have a higher chance of preserving information from radiation events, as a lower ratio of stabiliser qubits is corrupted.
}


\subsection{Effect of radiation fault position}
\label{rad_fault_pos}
In this analysis, we characterise the detection capabilities of the REI subroutine.
We test both the incidence of detection of a high energy event and the ratio of stabiliser qubits affected by the event.

In Figure \ref{rad_aoe}, we compare the radiation detection efficacy on the Rotated Surface code of distance $15$, drawing statistics for three separate radiation fault loci, highlighted in red in the left subplot.
We identify these injection points as \textit{Central}, \textit{North-West} and \textit{West}.
For all of them, the radiation event is injected at time $0$ \textit{ms} and lasts for $1$ \textit{ms}.
We want to highlight that the capabilities of REI are not affected by the fault location and even if corner qubits are affected we can still accurately identify the fault.

The middle subplot represents the evolution over time of the incidence of detection of a radiation event.
The REI subroutine immediately spots the presence of a radiation event, with a detection rate that sharply falls off towards the end of the event at $1 \ ms$.
The North-West injection point stops being detected about $50$ $\mu s$ earlier than the Central one, and the West injection point stops being detected about $100$ $\mu s$ before the Central one.
From this, it follows that the detection rate of the Central event, which affects a larger portion of qubits, is detected as active for a longer period of time.

In the rightmost plot we correlate the position of the radiation locus with the ratio of affected stabiliser qubits.
Since the intensity of the radiation event over the quantum chip's surface area dampens with inverse square proportionality, the further a locus of radiation is from the centre of the QEC code, the lower number of qubits, and thus stabilisers, are going to be affected.
We can notice this trend as the Central injection point retains the largest number of affected stabiliser qubits over time, whilst the West injection point retains the fewest.
Given that every location on the quantum chip is equally likely to be hit, we can state that the rate of affected stabiliser qubits is upper bounded in the worst case by a radiation event hitting the centre of the surface code.
However, a lower bound can not be identified, as the high energy event may impact a set of coordinates that lie outside of the area belonging to the surface code.
In this latter case, some of the stabiliser qubits might still be affected.

\obs{
The temporal persistence of a radiation event is correlated to the affected area in the surface code.
Peripheral faults induce transients that last for less time than those at the code's centre. 
}

\begin{figure*}[!ht]
    \centering
    \includegraphics[width=.9\linewidth]{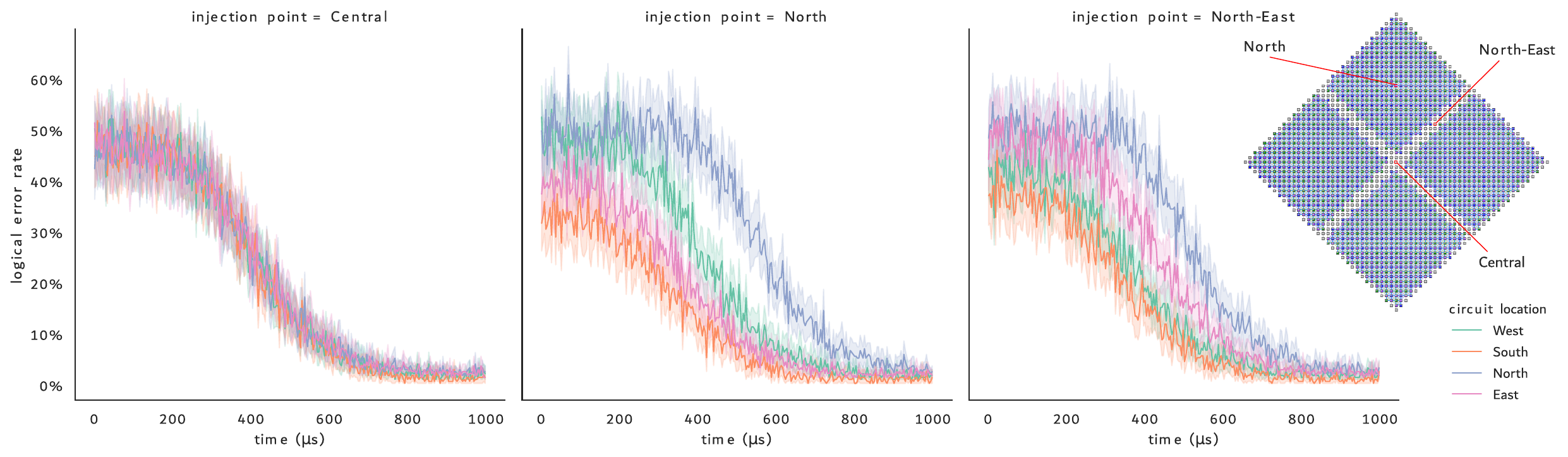}
        \caption{Multi-code logical error. We show three independent loci of radiation (right, outlined in red), affecting a quantum chip hosting four Rotated Surface codes of distance 15. The faults last for 1 $ms$, drawing statistics over 256 shots.}
    \label{multi_code_plot}
    
\end{figure*}

\begin{figure}[!t]
    \centering
    \includegraphics[width=.84\linewidth]{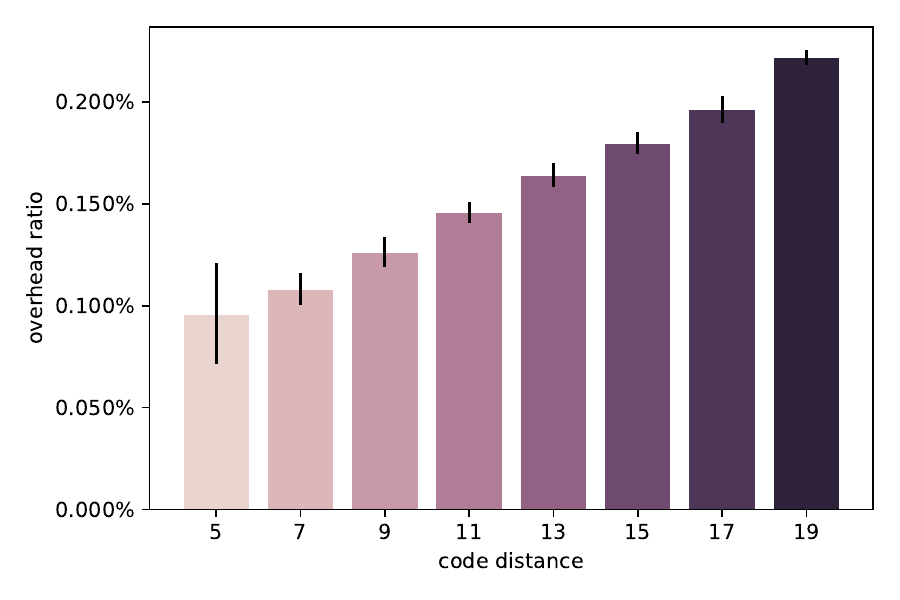}
        \caption{\textbf{Overhead ratio of radiation detection and of the MWPM decoder}. We plot the average overhead ratio on the Z-basis pass of the Rotated Surface code for growing code distance. Black bars represent the standard deviation.}
    \label{time_performance_detection}
    
\end{figure}

\subsection{Radiation detection complexity and time performance}
The scope of this analysis is to understand the impact of the radiation detection routine in the context of the constraints associated to QEC decoding in modern quantum computers.
Most importantly, we want to prove that the REI subroutine is efficient enough in terms of time to solution to be possibly integrated in existing decoders, as we will later show in Section \ref{decoder_performance_comparison}.
For this task, we once consider the Rotated Surface code in the Z basis, at distances ranging from 5 to 19.

In Figure \ref{time_performance_detection}, we measure the mean overhead required by an optimised C implementation of the REI subroutine compare to the time to solution of the Minimum Weight Perfect Matching decoder, which is the fastest in terms of time to solution between the decoders reported in Section \ref{other_decoders}.
For the sake of space, we did not report the relative overhead with respect to the other Tanner graph based decoders.
We considered a set of random stabiliser syndromes for both the REI subroutine and the MWPM decoder.
The average overhead induced by the subroutine ranges between $0.1\%$ for the distance three Rotated Surface code, to no more than $0.2\%$ for the same code at distance 19, despite the quadratic increase in the number of stabiliser qubits.
The REI subroutine requires from three to two orders of magnitude less time than the MWPM decoder to provide an output, thanks to early exit conditions that trigger in absence of a radiation event.
It is thus reasonable to assume the insertion of the REI subroutine as a component of future QEC decoders.

\obs{
Without the need for vectorisation, the sequential implementation of the REI subroutine introduces minimal overhead with respect to the decoding step.
}

Most of the complexity of the REI subroutine stems from the computation of the euclidean distance between two vectors of coordinates of size $n$, which add up to a complexity of $O(n^2)$ in the number of comparisons, whilst the remainder of the operations have complexity $O(n)$, where in the worst case $n=(d^2-1)$, with $d$ being the distance of the surface code.
This scales favourably with respect to the $O(d^6 log(d))$ complexity of the MWPM decoder for a code of distance $d$ \cite{pymatching}. 
Since the highest complexity operations of the REI subroutine are easily vectorisable and parallelisable, the time to solution would reach similar acceleration as that of hardware based decoders in use in real-world quantum computers.

\begin{figure*}[!ht]
    \centering
    \includegraphics[width=\linewidth]{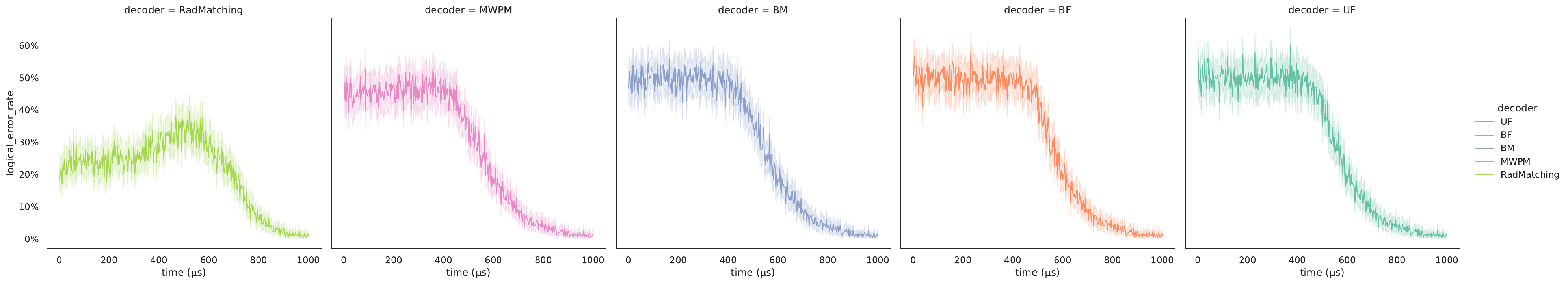}
        \caption{\textbf{Decoder performance comparison}. We considered the Rotated Surface code of distance 9 with 9 repetitions per round of correction. The RadMatching decoder is compared with other graph based decoders: Minimum Weight Perfect Matching (MWPM), BeliefMatching (BM), BeliefFind (BF) and UnionFind (UF). Decoders are ordered from best to worst going left to right. Statistics are drawn over 384 shots.}
    \label{decoder_comparison_fig}
    
\end{figure*}

\subsection{Multi-code logical error correlation}
\label{multi_code_analysis}
Recent years have seen a surge in the number of physical qubits embedded in a single superconducting chip.
There is reason to believe that this trend will continue, and that multiple QEC codes may be placed onto the same quantum chip.
In this analysis, we extrapolate the logical error rates of a future large-scale superconducting quantum chip.
We consider an expanded chip structure, with the same repeating pattern as the one presented in Figure \ref{chip_topology}, hosting four distance 15 Rotated Surface codes, placed respectively in North, South, East and West quadrants obtained by cutting the quantum chip with two orthogonal diagonal lines.
We report the performance of the Minimum Weight Perfect Matching decoder with respect to three independent loci for the radiation fault in separate simulations.
Similarly to what had been presented in Section \ref{rad_fault_pos}, we identify one \textit{Central} fault position at the centre of the quantum chip, equidistant to all surface codes, one fault position at the halfway \textit{North-East} point between two surface codes, and one fault position centred onto the \textit{North} surface code.

In Figure \ref{multi_code_plot} we plot the post-decoding logical error rate of the four surface codes as a function of time, considering a radiation fault lasting for $1$ $ms$.

\obs{
Surface codes which are equidistant from a radiation event will experience correlated logical error spikes of similar intensity.
}

In the leftmost subplot, we see how all surface codes are equally affected by the Central radiation fault, with logical rates that closely match one another.
This behaviour is expected, since they are all equidistant from the locus of radiation.

Conversely, when considering the North injection point in the middle subplot, the North surface code is affected to a higher degree by the radiation fault, reaching a peak of more than $53\%$ logical error rate for more than half of the total fault's duration, whilst the second closest is the West surface code, with a peak at about $45\%$, closely followed by the East code with a similar peak at $41\%$, while the furthest surface code, in the South quadrant reaches a lower peak of $34\%$.
Notably, the overall persistence of the peak is longer for the North code, which is affected the most, while in further away codes the logical error converges to zero faster.
In the leftmost plot, we consider the North-East locus of radiation, where both the North and the East codes reach a peak logic error rate of about $50\%$, while the West and South code reach lower peaks of about $40\%$.
Physical distance from the impact point is thus insufficient to prevent multi-code logical errors.
We can thus expect bundles of proximal logical qubits to be subject to correlated logical error spikes.

\obs{
Radiation events reaching far away surface codes can still overcome their error correction threshold, although to a lesser degree than closer events.
}

\subsection{Radiation aware decoding}
\label{decoder_performance_comparison}

We conjecture that, by using information from the REI subroutine (fault detection, fault location identification, and area affected by radiation), we can partly mitigate the effects of radiation on the Rotated Surface code during the decoding phase.
Intuitively, since the stabilisers are physically interleaved with data qubits, and radiation spreads in space, we expect that if a group of stabilisers has been corrupted, the data qubits inside that area have been affected by radiation as well. 
Once the area of effect of the radiation event has been identified, we apply a bitwise inversion to all of the stabiliser measurements in the current syndrome that were hosted on physical qubits inside the affected area, mapping \textit{true} values to \textit{false}, and vice versa. 
If a stabiliser measurement belongs to a qubit outside the radiation fault, it is left unmodified.
This processed syndrome is then fed to the Minumum Weight Perfect Matching decoder, which will provide an output prediction for the state of the QEC code, to be compared with the rest of the decoders, as such we label it the \textit{RadMatching} decoder.  

To measure if RadMatching can be effectively used to compensate for radiation-induced events, we compare the logical error rate of the Rotated Surface code (distance 9 and with 9 repetitions per error correction round)  \textit{after} the decoding process adopting RadMatching versus adopting the Tanner graph based decoders presented in Section~\ref{other_decoders}. 
We consider a single radiation fault, lasting for $1$ $ms$, injected at the centre of the quantum chip, where radiation's effects are most detrimental. 
We compare the performance of all decoders onto the same temporal sequence of syndrome measurements.

In Figure~\ref{decoder_comparison_fig}, we plot the logical error of the Rotated Surface code over the time window of the radiation fault.
The performance of our RadMatching implementation is shown on the leftmost subplot, using the REI subroutine together with the MWPM decoder.
Our approach improves onto the logical error of the non radiation-aware MWPM decoder by upwards of $25\%$ at the beginning of the radiation event, when the fault's intensity is maximal.

\obs{
Identification of radiation's area-of-effect provides vital information to the RadMatching decoder, lowering the overall logical error rate of the Rotated Surface code.
}

All the other non-radiation aware graph based decoders considered in this analysis show marginally worse performance than the MWPM decoder, with the BM decoder reaching a peak of $50\%$, the BF decoder reaching a peak of $52\%$ and the UF decoder reaching again a peak of $50\%$.
All of the four non radiation-aware decoders maintain their respective logical error peak from more than half of the total duration of the fault.
On the other hand, the RadMatching decoder shows a relatively stable logical error rate of about $25\%$ in the first third of the radiation event's duration, albeit reaching a logical error of about $35 \%$ at the halfway point, which is slightly higher than that at the beginning of the time window. 
This slight performance degradation (which still shows an improvement over existing decoders) is due to the lower radiation detection rate after the halfway point, since the intensity of the radiation fault gets smaller over time, as shown in Section \ref{code_distance_detection}.
In the last portion of the time evolution, all the considered decoders converge to their respective nominal \textit{radiation-free} logical error rates.

\obs{
Tanner-graph based decoders which are not radiation aware can not keep up with the error rates imposed by radiation events.
}

\section{Conclusions and future works}
\label{conclusions_and_future_works}
In this work, we have considered how to model the impact of a high energy event onto a superconducting quantum chip, then we ran simulations on multiple configurations of the Rotated Surface code.

Reaching back to the introductory research questions of this article, we tackled \textbf{RQ1} by introducing the novel REI subroutine, which makes it possible to identify the incidence and the area-of-effect of radiation events at runtime, with a time to solution which is just a fraction of that of decoding.
We underlined important observations relative to \textbf{RQ2}, stating that larger surface codes are inherently more likely to preserve stabiliser information in the event of radiation faults, and that the most detrimental location for a fault to happen is right at the centre of the surface code's area, with longer lasting transient effects with respect to peripheral positions.
Moreover, we deepened our answer to \textbf{RQ2} by considering the post-decoding logical error rates of a set of Rotated Surface codes on a shared quantum chip, noticing how their logical error rate is correlated to their respective distance from the impact point, hinting at how radiation-induced faults might propagate across logical qubits.
Furthermore, we compared a set of Tanner-graph based decoders, highlighting that standard decoding techniques are not sufficient for dealing with radiation events.
At last, we assess \textbf{RQ3} by measuring the performance of the RadMatching decoder, a MWPM decoder made radiation-aware by the REI subroutine, showing an average reduction of the logical error rate by about $25\%$ with respect to other approaches.

This work stresses the urgency of of finding new and alternative solutions to the issue imposed by high energy events in superconducting quantum computers.
Despite the promising results, there is still plenty of research to be done, regarding extensions and generalisations of radiation-aware decoding techniques, the verification of new QEC code classes, and the analysis of extended prototypes for quantum memories.


\printbibliography

@String{Computing = "Computing" }

@String{Computer = "{IEEE} Computer" }

@String{Springer = "Springer-Verlag" }

@article{hot_qubit,
  title = {Superconducting Qubits above 20 GHz Operating over 200 mK},
  author = {Anferov, Alexander and Harvey, Shannon P. and Wan, Fanghui and Simon, Jonathan and Schuster, David I.},
  journal = {PRX Quantum},
  volume = {5},
  issue = {3},
  pages = {030347},
  numpages = {19},
  year = {2024},
  month = {9},
  publisher = {American Physical Society},
  doi = {10.1103/PRXQuantum.5.030347},
  url = {https://link.aps.org/doi/10.1103/PRXQuantum.5.030347}
}

@ArtifactSoftware{R,
    title = {R: A Language and Environment for Statistical Computing},
    author = {{R Core Team}},
    organization = {R Foundation for Statistical Computing},
    address = {Vienna, Austria},
    year = {2019},
    url = {https://www.R-project.org/},
}

@article{mcewen2022resolving,
  title    = {Resolving catastrophic error bursts from cosmic rays in large arrays of superconducting qubits},
  volume   = {18},
  issn     = {1745-2481},
  url      = {https://doi.org/10.1038/s41567-021-01432-8},
  doi      = {10.1038/s41567-021-01432-8},
  number   = {1},
  journal  = {Nature Physics},
  author   = {McEwen, Matt and Faoro, Lara and Arya, Kunal and Dunsworth, Andrew and Huang, Trent and Kim, Seon and Burkett, Brian and Fowler, Austin and Arute, Frank and Bardin, Joseph C. and Bengtsson, Andreas and Bilmes, Alexander and Buckley, Bob B. and Bushnell, Nicholas and Chen, Zijun and Collins, Roberto and Demura, Sean and Derk, Alan R. and Erickson, Catherine and Giustina, Marissa and Harrington, Sean D. and Hong, Sabrina and Jeffrey, Evan and Kelly, Julian and Klimov, Paul V. and Kostritsa, Fedor and Laptev, Pavel and Locharla, Aditya and Mi, Xiao and Miao, Kevin C. and Montazeri, Shirin and Mutus, Josh and Naaman, Ofer and Neeley, Matthew and Neill, Charles and Opremcak, Alex and Quintana, Chris and Redd, Nicholas and Roushan, Pedram and Sank, Daniel and Satzinger, Kevin J. and Shvarts, Vladimir and White, Theodore and Yao, Z. Jamie and Yeh, Ping and Yoo, Juhwan and Chen, Yu and Smelyanskiy, Vadim and Martinis, John M. and Neven, Hartmut and Megrant, Anthony and Ioffe, Lev and Barends, Rami},
  month    = jan,
  year     = {2022},
  pages    = {107--111}
}

@article{Baumann2005,
  author   = {Robert Baumann},
  journal  = {IEEE Design Test of Computers},
  title    = {Soft errors in advanced computer systems},
  year     = {2005},
  volume   = {22},
  number   = {3},
  pages    = {258-266},
  doi      = {10.1109/MDT.2005.69},
  issn     = {0740-7475},
      month    = {5}
}

@techreport{Jedec2006,
  author      = {JEDEC},
  institution = {JEDEC Standard},
  number      = {JESD89A},
  title       = {{Measurement and Reporting of Alpha Particle and Terrestrial Cosmic Ray-Induced Soft Errors in Semiconductor Devices}},
  year        = {2006}
}

@article{Vepsalainen2021,
  author        = {Veps{\"a}l{\"a}inen, Antti P. and Karamlou, Amir H. and Orrell, John L. and Dogra, Akshunna S. and Loer, Ben and Vasconcelos, Francisca and Kim, David K. and Melville, Alexander J. and Niedzielski, Bethany M. and Yoder, Jonilyn L. and Gustavsson, Simon and Formaggio, Joseph A. and VanDevender, Brent A. and Oliver, William D.},
  da            = {2020/08/01},
  doi           = {10.1038/s41586-020-2619-8},
  id            = {Veps{\"a}l{\"a}inen2020},
  journal       = {Nature},
  number        = {7822},
  pages         = {551--556},
  title         = {Impact of ionizing radiation on superconducting qubit coherence},
  ty            = {JOUR},
  url           = {https://doi.org/10.1038/s41586-020-2619-8},
  volume        = {584},
  year          = {2020}
}

@article{Wilen2021,
  author={Wilen, Christopher D and Abdullah, S and Kurinsky, NA and Stanford, C and Cardani, L and d’Imperio, G and Tomei, C and Faoro, L and Ioffe, LB and Liu, CH and others},
  da            = {2021/06/01},
  doi           = {10.1038/s41586-021-03557-5},
  id            = {Wilen2021},
  journal       = {Nature},
  number        = {7863},
  pages         = {369--373},
  title         = {Correlated charge noise and relaxation errors in superconducting qubits},
  ty            = {JOUR},
  url           = {https://doi.org/10.1038/s41586-021-03557-5},
  volume        = {594},
  year          = {2021}
}

@article{Martinis2021,
  author        = {Martinis, John M. },
  da            = {2021/06/03},
  doi           = {10.1038/s41534-021-00431-0},
  id            = {Martinis2021},
  journal       = {npj Quantum Information},
  number        = {1},
  pages         = {90},
  title         = {Saving superconducting quantum processors from decay and correlated errors generated by gamma and cosmic rays},
  ty            = {JOUR},
  url           = {https://doi.org/10.1038/s41534-021-00431-0},
  volume        = {7},
  year          = {2021}
}

@article{Cardani2021,
  title={Reducing the impact of radioactivity on quantum circuits in a deep-underground facility},
  author={Cardani, Laura and Valenti, Francesco and Casali, Nicola and Catelani, Gianluigi and Charpentier, Thibault and Clemenza, Massimiliano and Colantoni, Ivan and Cruciani, Angelo and D’Imperio, G and Gironi, Luca and others},
  journal={Nature communications},
  volume={12},
  number={1},
  pages={2733},
  year={2021},
  publisher={Nature Publishing Group UK London}
}

@article{Preskill_2018,
  title     = {Quantum Computing in the NISQ era and beyond},
  volume    = {2},
  issn      = {2521-327X},
  url       = {http://dx.doi.org/10.22331/q-2018-08-06-79},
  doi       = {10.22331/q-2018-08-06-79},
  journal   = {Quantum},
  publisher = {Verein zur Forderung des Open Access Publizierens in den Quantenwissenschaften},
  author    = {Preskill, John},
  year      = {2018},
  month     = {8},
  pages     = {79}
}

@article{QuFI,
  title   = {QuFI: a Quantum Fault Injector to Measure the Reliability of Qubits and Quantum Circuits},
  author  = {Oliveira, Daniel and Giusto, Edoardo and Dri, Emanuele and Casciola, Nadir and Baheri, Betis and Guan, Qiang and Montrucchio, Bartolomeo and Rech, Paolo},
  journal = {arXiv preprint arXiv:2203.07183},
  year    = {2022}
}

@misc{repo,
  title        = {Work repository to be disclosed after the review process},
  howpublished = {\url{https://github.com/}},
  author       = {Anonymous Author},
  year         = {2025},
}

@article{Cardani2023,
  author   = {Cardani, L.
              and Colantoni, I.
              and Cruciani, A.
              and De Dominicis, F.
              and D'Imperio, G.
              and Laubenstein, M.
              and Mariani, A.
              and Pagnanini, L.
              and Pirro, S.
              and Tomei, C.
              and Casali, N.
              and Ferroni, F.
              and Frolov, D.
              and Gironi, L.
              and Grassellino, A.
              and Junker, M.
              and Kopas, C.
              and Lachman, E.
              and McRae, C. R. H.
              and Mutus, J.
              and Nastasi, M.
              and Pappas, D. P.
              and Pilipenko, R.
              and Sisti, M.
              and Pettinacci, V.
              and Romanenko, A.
              and Van Zanten, D.
              and Vignati, M.
              and Withrow, J. D.
              and Zhelev, N. Z.},
  title    = {Disentangling the sources of ionizing radiation in superconducting qubits},
  journal  = {The European Physical Journal C},
  year     = {2023},
  month    = {1},
  day      = {31},
  volume   = {83},
  number   = {1},
  pages    = {94},
  issn     = {1434-6052},
  doi      = {10.1140/epjc/s10052-023-11199-2},
  url      = {https://doi.org/10.1140/epjc/s10052-023-11199-2}
}

@article{Junger2025suspendedqubits,
    author = {Jünger, Christian and Chistolini, Trevor and Nguyen, Long B. and Kim, Hyunseong and Chen, Larry and Ersevim, Thomas and Livingston, William and Koolstra, Gerwin and Santiago, David I. and Siddiqi, Irfan},
    title = {Implementation of scalable suspended superinductors},
    journal = {Applied Physics Letters},
    volume = {126},
    number = {4},
    pages = {044003},
    year = {2025},
    month = {01},
    issn = {0003-6951},
    doi = {10.1063/5.0250341},
    url = {https://doi.org/10.1063/5.0250341},
    eprint = {https://pubs.aip.org/aip/apl/article-pdf/doi/10.1063/5.0250341/20367496/044003\_1\_5.0250341.pdf},
}

@article{Bombin2007rotatedsurfacecode,
   title={Optimal resources for topological two-dimensional stabilizer codes: Comparative study},
   volume={76},
   ISSN={1094-1622},
   url={http://dx.doi.org/10.1103/PhysRevA.76.012305},
   DOI={10.1103/physreva.76.012305},
   number={1},
   journal={Physical Review A},
   publisher={American Physical Society (APS)},
   author={Bombin, H. and Martin-Delgado, M. A.},
   year={2007},
   month=jul }

@inproceedings{Kovalev2012rotatedsurfacecode,
   title={Improved quantum hypergraph-product LDPC codes},
   url={http://dx.doi.org/10.1109/ISIT.2012.6284206},
   DOI={10.1109/isit.2012.6284206},
   booktitle={2012 IEEE International Symposium on Information Theory Proceedings},
   publisher={IEEE},
   author={Kovalev, Alexey A. and Pryadko, Leonid P.},
   year={2012},
   month=jul, pages={348–352} }

@misc{gottesman1997qec,
      title={Stabilizer Codes and Quantum Error Correction}, 
      author={Daniel Gottesman},
      year={1997},
      eprint={quant-ph/9705052},
      archivePrefix={arXiv},
      primaryClass={quant-ph},
      url={https://arxiv.org/abs/quant-ph/9705052}, 
}

@INPROCEEDINGS{Chatterjee2023qec_for_dummies,
  author={Chatterjee, Avimita and Phalak, Koustubh and Ghosh, Swaroop},
  booktitle={2023 IEEE International Conference on Quantum Computing and Engineering (QCE)}, 
  title={Quantum Error Correction For Dummies}, 
  year={2023},
  volume={01},
  number={},
  pages={70-81},
  doi={10.1109/QCE57702.2023.00017}}

@Article{Acharya2023,
author={Acharya, Rajeev
and Aleiner, Igor
and Allen, Richard
and Andersen, Trond I.
and Ansmann, Markus
and Arute, Frank
and Arya, Kunal
and Asfaw, Abraham
and Atalaya, Juan
and Babbush, Ryan
and Bacon, Dave
and Bardin, Joseph C.
and Basso, Joao
and Bengtsson, Andreas
and Boixo, Sergio
and Bortoli, Gina
and Bourassa, Alexandre
and Bovaird, Jenna
and Brill, Leon
and Broughton, Michael
and Buckley, Bob B.
and Buell, David A.
and Burger, Tim
and Burkett, Brian
and Bushnell, Nicholas
and Chen, Yu
and Chen, Zijun
and Chiaro, Ben
and Cogan, Josh
and Collins, Roberto
and Conner, Paul
and Courtney, William
and Crook, Alexander L.
and Curtin, Ben
and Debroy, Dripto M.
and Del Toro Barba, Alexander
and Demura, Sean
and Dunsworth, Andrew
and Eppens, Daniel
and Erickson, Catherine
and Faoro, Lara
and Farhi, Edward
and Fatemi, Reza
and Flores Burgos, Leslie
and Forati, Ebrahim
and Fowler, Austin G.
and Foxen, Brooks
and Giang, William
and Gidney, Craig
and Gilboa, Dar
and Giustina, Marissa
and Grajales Dau, Alejandro
and Gross, Jonathan A.
and Habegger, Steve
and Hamilton, Michael C.
and Harrigan, Matthew P.
and Harrington, Sean D.
and Higgott, Oscar
and Hilton, Jeremy
and Hoffmann, Markus
and Hong, Sabrina
and Huang, Trent
and Huff, Ashley
and Huggins, William J.
and Ioffe, Lev B.
and Isakov, Sergei V.
and Iveland, Justin
and Jeffrey, Evan
and Jiang, Zhang
and Jones, Cody
and Juhas, Pavol
and Kafri, Dvir
and Kechedzhi, Kostyantyn
and Kelly, Julian
and Khattar, Tanuj
and Khezri, Mostafa
and Kieferov{\'a}, M{\'a}ria
and Kim, Seon
and Kitaev, Alexei
and Klimov, Paul V.
and Klots, Andrey R.
and Korotkov, Alexander N.
and Kostritsa, Fedor
and Kreikebaum, John Mark
and Landhuis, David
and Laptev, Pavel
and Lau, Kim-Ming
and Laws, Lily
and Lee, Joonho
and Lee, Kenny
and Lester, Brian J.
and Lill, Alexander
and Liu, Wayne
and Locharla, Aditya
and Lucero, Erik
and Malone, Fionn D.
and Marshall, Jeffrey
and Martin, Orion
and McClean, Jarrod R.
and McCourt, Trevor
and McEwen, Matt
and Megrant, Anthony
and Meurer Costa, Bernardo
and Mi, Xiao
and Miao, Kevin C.
and Mohseni, Masoud
and Montazeri, Shirin
and Morvan, Alexis
and Mount, Emily
and Mruczkiewicz, Wojciech
and Naaman, Ofer
and Neeley, Matthew
and Neill, Charles
and Nersisyan, Ani
and Neven, Hartmut
and Newman, Michael
and Ng, Jiun How
and Nguyen, Anthony
and Nguyen, Murray
and Niu, Murphy Yuezhen
and O'Brien, Thomas E.
and Opremcak, Alex
and Platt, John
and Petukhov, Andre
and Potter, Rebecca
and Pryadko, Leonid P.
and Quintana, Chris
and Roushan, Pedram
and Rubin, Nicholas C.
and Saei, Negar
and Sank, Daniel
and Sankaragomathi, Kannan
and Satzinger, Kevin J.
and Schurkus, Henry F.
and Schuster, Christopher
and Shearn, Michael J.
and Shorter, Aaron
and Shvarts, Vladimir
and Skruzny, Jindra
and Smelyanskiy, Vadim
and Smith, W. Clarke
and Sterling, George
and Strain, Doug
and Szalay, Marco
and Torres, Alfredo
and Vidal, Guifre
and Villalonga, Benjamin
and Vollgraff Heidweiller, Catherine
and White, Theodore
and Xing, Cheng
and Yao, Z. Jamie
and Yeh, Ping
and Yoo, Juhwan
and Young, Grayson
and Zalcman, Adam
and Zhang, Yaxing
and Zhu, Ningfeng
and AI, Google Quantum},
title={Suppressing quantum errors by scaling a surface code logical qubit},
journal={Nature},
year={2023},
month={2},
day={01},
volume={614},
number={7949},
pages={676-681},
issn={1476-4687},
doi={10.1038/s41586-022-05434-1},
url={https://doi.org/10.1038/s41586-022-05434-1}
}

@misc{harrington2024synchronous,
      title={Synchronous Detection of Cosmic Rays and Correlated Errors in Superconducting Qubit Arrays}, 
      author={Patrick M. Harrington and Mingyu Li and Max Hays and Wouter Van De Pontseele and Daniel Mayer and H. Douglas Pinckney and Felipe Contipelli and Michael Gingras and Bethany M. Niedzielski and Hannah Stickler and Jonilyn L. Yoder and Mollie E. Schwartz and Jeffrey A. Grover and Kyle Serniak and William D. Oliver and Joseph A. Formaggio},
      year={2024},
      eprint={2402.03208},
      archivePrefix={arXiv},
      primaryClass={quant-ph}
}

@article{Loer2024underground,
doi = {10.1088/1748-0221/19/09/P09001},
url = {https://dx.doi.org/10.1088/1748-0221/19/09/P09001},
year = {2024},
month = {9},
publisher = {IOP Publishing},
volume = {19},
number = {09},
pages = {P09001},
author = {Loer, B. and Harrington, P.M. and Archambault, B. and Fuller, E. and Pierson, B. and Arnquist, I.J. and Harouaka, K. and Schlieder, T.D. and Kim, D.K. and Melville, A.J. and Niedzielski, B.M. and Yoder, J.L. and Serniak, K. and Oliver, W.D. and Orrell, J.L. and Bunker, R. and VanDevender, B.A. and Warner, M.},
title = {Abatement of ionizing radiation for superconducting quantum devices},
journal = {Journal of Instrumentation}
}

@article{Georgopoulos2021,
  title = {Modeling and simulating the noisy behavior of near-term quantum computers},
  author = {Georgopoulos, Konstantinos and Emary, Clive and Zuliani, Paolo},
  journal = {Phys. Rev. A},
  volume = {104},
  issue = {6},
  pages = {062432},
  numpages = {14},
  year = {2021},
  month = {12},
  publisher = {American Physical Society},
  doi = {10.1103/PhysRevA.104.062432},
  url = {https://link.aps.org/doi/10.1103/PhysRevA.104.062432}
}

@Article{Wootters1982nocloning,
author={Wootters, W. K.
and Zurek, W. H.},
title={A single quantum cannot be cloned},
journal={Nature},
year={1982},
month={10},
day={01},
volume={299},
number={5886},
pages={802-803},
issn={1476-4687},
doi={10.1038/299802a0},
url={https://doi.org/10.1038/299802a0}
}

@INPROCEEDINGS{vallero2024efficacy,
  author={Vallero, Marzio and Casagranda, Gioele and Vella, Flavio and Rech, Paolo},
  booktitle={SC24: International Conference for High Performance Computing, Networking, Storage and Analysis}, 
  title={On the Efficacy of Surface Codes in Compensating for Radiation Events in Superconducting Devices}, 
  year={2024},
  volume={},
  number={},
  pages={1-15},
  doi={10.1109/SC41406.2024.00075}}

@ARTICLE{casagranda2025tns,
  author={Casagranda, Gioele and Vallero, Marzio and Vella, Flavio and Rech, Paolo},
  journal={IEEE Transactions on Nuclear Science}, 
  title={Understanding the Contributions of Terrestrial Radiation Sources to Error Rates in Quantum Devices}, 
  year={2025},
  volume={},
  number={},
  pages={1-1},
  doi={10.1109/TNS.2025.3537036}}

@article{gidney2021stim,
  doi = {10.22331/q-2021-07-06-497},
  url = {https://doi.org/10.22331/q-2021-07-06-497},
  title = {Stim: a fast stabilizer circuit simulator},
  author = {Gidney, Craig},
  journal = {{Quantum}},
  issn = {2521-327X},
  publisher = {{Verein zur F{\"{o}}rderung des Open Access Publizierens in den Quantenwissenschaften}},
  volume = {5},
  pages = {497},
  month = jul,
  year = {2021}
}

@article{pymatching,
author = {Higgott, Oscar},
title = {PyMatching: A Python Package for Decoding Quantum Codes with Minimum-Weight Perfect Matching},
year = {2022},
issue_date = {September 2022},
publisher = {Association for Computing Machinery},
address = {New York, NY, USA},
volume = {3},
number = {3},
url = {https://doi.org/10.1145/3505637},
doi = {10.1145/3505637},
journal = {ACM Transactions on Quantum Computing},
month = jun,
articleno = {16},
numpages = {16}
}

@misc{Roffe_LDPC_Python_tools_2022,
author = {Roffe, Joschka},
title = {{LDPC: Python tools for low density parity check codes}},
url = {https://pypi.org/project/ldpc/},
year = {2022}
}

@article{Delfosse_2021,
   title={Almost-linear time decoding algorithm for topological codes},
   volume={5},
   ISSN={2521-327X},
   url={http://dx.doi.org/10.22331/q-2021-12-02-595},
   DOI={10.22331/q-2021-12-02-595},
   journal={Quantum},
   publisher={Verein zur Forderung des Open Access Publizierens in den Quantenwissenschaften},
   author={Delfosse, Nicolas and Nickerson, Naomi H.},
   year={2021},
   month=dec, pages={595} }

@article{delfosse2021unionfinddecoderquantumldpc,
  author={Delfosse, Nicolas and Londe, Vivien and Beverland, Michael E.},
  journal={IEEE Transactions on Information Theory}, 
  title={Toward a Union-Find Decoder for Quantum LDPC Codes}, 
  year={2022},
  volume={68},
  number={5},
  pages={3187-3199},
  doi={10.1109/TIT.2022.3143452}
}

@misc{higgott2023improveddecodingcircuitnoise,
      title={Improved decoding of circuit noise and fragile boundaries of tailored surface codes}, 
      author={Oscar Higgott and Thomas C. Bohdanowicz and Aleksander Kubica and Steven T. Flammia and Earl T. Campbell},
      year={2023},
      eprint={2203.04948},
      archivePrefix={arXiv},
      primaryClass={quant-ph},
      url={https://arxiv.org/abs/2203.04948}, 
}

@Article{Acharya2025,
author={Acharya, Rajeev
and Abanin, Dmitry A.
and Aghababaie-Beni, Laleh
and Aleiner, Igor
and Andersen, Trond I.
and Ansmann, Markus
and Arute, Frank
and Arya, Kunal
and Asfaw, Abraham
and Astrakhantsev, Nikita
and Atalaya, Juan
and Babbush, Ryan
and Bacon, Dave
and Ballard, Brian
and Bardin, Joseph C.
and Bausch, Johannes
and Bengtsson, Andreas
and Bilmes, Alexander
and Blackwell, Sam
and Boixo, Sergio
and Bortoli, Gina
and Bourassa, Alexandre
and Bovaird, Jenna
and Brill, Leon
and Broughton, Michael
and Browne, David A.
and Buchea, Brett
and Buckley, Bob B.
and Buell, David A.
and Burger, Tim
and Burkett, Brian
and Bushnell, Nicholas
and Cabrera, Anthony
and Campero, Juan
and Chang, Hung-Shen
and Chen, Yu
and Chen, Zijun
and Chiaro, Ben
and Chik, Desmond
and Chou, Charina
and Claes, Jahan
and Cleland, Agnetta Y.
and Cogan, Josh
and Collins, Roberto
and Conner, Paul
and Courtney, William
and Crook, Alexander L.
and Curtin, Ben
and Das, Sayan
and Davies, Alex
and De Lorenzo, Laura
and Debroy, Dripto M.
and Demura, Sean
and Devoret, Michel
and Di Paolo, Agustin
and Donohoe, Paul
and Drozdov, Ilya
and Dunsworth, Andrew
and Earle, Clint
and Edlich, Thomas
and Eickbusch, Alec
and Elbag, Aviv Moshe
and Elzouka, Mahmoud
and Erickson, Catherine
and Faoro, Lara
and Farhi, Edward
and Ferreira, Vinicius S.
and Burgos, Leslie Flores
and Forati, Ebrahim
and Fowler, Austin G.
and Foxen, Brooks
and Ganjam, Suhas
and Garcia, Gonzalo
and Gasca, Robert
and Genois, {\'E}lie
and Giang, William
and Gidney, Craig
and Gilboa, Dar
and Gosula, Raja
and Dau, Alejandro Grajales
and Graumann, Dietrich
and Greene, Alex
and Gross, Jonathan A.
and Habegger, Steve
and Hall, John
and Hamilton, Michael C.
and Hansen, Monica
and Harrigan, Matthew P.
and Harrington, Sean D.
and Heras, Francisco J. H.
and Heslin, Stephen
and Heu, Paula
and Higgott, Oscar
and Hill, Gordon
and Hilton, Jeremy
and Holland, George
and Hong, Sabrina
and Huang, Hsin-Yuan
and Huff, Ashley
and Huggins, William J.
and Ioffe, Lev B.
and Isakov, Sergei V.
and Iveland, Justin
and Jeffrey, Evan
and Jiang, Zhang
and Jones, Cody
and Jordan, Stephen
and Joshi, Chaitali
and Juhas, Pavol
and Kafri, Dvir
and Kang, Hui
and Karamlou, Amir H.
and Kechedzhi, Kostyantyn
and Kelly, Julian
and Khaire, Trupti
and Khattar, Tanuj
and Khezri, Mostafa
and Kim, Seon
and Klimov, Paul V.
and Klots, Andrey R.
and Kobrin, Bryce
and Kohli, Pushmeet
and Korotkov, Alexander N.
and Kostritsa, Fedor
and Kothari, Robin
and Kozlovskii, Borislav
and Kreikebaum, John Mark
and Kurilovich, Vladislav D.
and Lacroix, Nathan
and Landhuis, David
and Lange-Dei, Tiano
and Langley, Brandon W.
and Laptev, Pavel
and Lau, Kim-Ming
and Le Guevel, Lo{\"i}ck
and Ledford, Justin
and Lee, Joonho
and Lee, Kenny
and Lensky, Yuri D.
and Leon, Shannon
and Lester, Brian J.
and Li, Wing Yan
and Li, Yin
and Lill, Alexander T.
and Liu, Wayne
and Livingston, William P.
and Locharla, Aditya
and Lucero, Erik
and Lundahl, Daniel
and Lunt, Aaron
and Madhuk, Sid
and Malone, Fionn D.
and Maloney, Ashley
and Mandr{\`a}, Salvatore
and Manyika, James
and Martin, Leigh S.
and Martin, Orion
and Martin, Steven
and Maxfield, Cameron
and McClean, Jarrod R.
and McEwen, Matt
and Meeks, Seneca
and Megrant, Anthony
and Mi, Xiao
and Miao, Kevin C.
and Mieszala, Amanda
and Molavi, Reza
and Molina, Sebastian
and Montazeri, Shirin
and Morvan, Alexis
and Movassagh, Ramis
and Mruczkiewicz, Wojciech
and Naaman, Ofer
and Neeley, Matthew
and Neill, Charles
and Nersisyan, Ani
and Neven, Hartmut
and Newman, Michael
and Ng, Jiun How
and Nguyen, Anthony
and Nguyen, Murray
and Ni, Chia-Hung
and Niu, Murphy Yuezhen
and O'Brien, Thomas E.
and Oliver, William D.
and Opremcak, Alex
and Ottosson, Kristoffer
and Petukhov, Andre
and Pizzuto, Alex
and Platt, John
and Potter, Rebecca
and Pritchard, Orion
and Pryadko, Leonid P.
and Quintana, Chris
and Ramachandran, Ganesh
and Reagor, Matthew J.
and Redding, John
and Rhodes, David M.
and Roberts, Gabrielle
and Rosenberg, Eliott
and Rosenfeld, Emma
and Roushan, Pedram
and Rubin, Nicholas C.
and Saei, Negar
and Sank, Daniel
and Sankaragomathi, Kannan
and Satzinger, Kevin J.
and Schurkus, Henry F.
and Schuster, Christopher
and Senior, Andrew W.
and Shearn, Michael J.
and Shorter, Aaron
and Shutty, Noah
and Shvarts, Vladimir
and Singh, Shraddha
and Sivak, Volodymyr
and Skruzny, Jindra
and Small, Spencer
and Smelyanskiy, Vadim
and Smith, W. Clarke
and Somma, Rolando D.
and Springer, Sofia
and Sterling, George
and Strain, Doug
and Suchard, Jordan
and Szasz, Aaron
and Sztein, Alex
and Thor, Douglas
and Torres, Alfredo
and Torunbalci, M. Mert
and Vaishnav, Abeer
and Vargas, Justin
and Vdovichev, Sergey
and Vidal, Guifre
and Villalonga, Benjamin
and Heidweiller, Catherine Vollgraff
and Waltman, Steven
and Wang, Shannon X.
and Ware, Brayden
and Weber, Kate
and Weidel, Travis
and White, Theodore
and Wong, Kristi
and Woo, Bryan W. K.
and Xing, Cheng
and Yao, Z. Jamie
and Yeh, Ping
and Ying, Bicheng
and Yoo, Juhwan
and Yosri, Noureldin
and Young, Grayson
and Zalcman, Adam
and Zhang, Yaxing
and Zhu, Ningfeng
and Zobrist, Nicholas
and AI, Google Quantum
and {Collaborators}},
title={Quantum error correction below the surface code threshold},
journal={Nature},
year={2025},
month={2},
day={01},
volume={638},
number={8052},
pages={920-926},
issn={1476-4687},
doi={10.1038/s41586-024-08449-y},
url={https://doi.org/10.1038/s41586-024-08449-y}
}

@article{Edmonds1965blossom, title={Paths, Trees, and Flowers}, volume={17}, DOI={10.4153/CJM-1965-045-4}, journal={Canadian Journal of Mathematics}, author={Edmonds, Jack}, year={1965}, pages={449–467}}

@Article{Bravyi2024GrossCode,
author={Bravyi, Sergey
and Cross, Andrew W.
and Gambetta, Jay M.
and Maslov, Dmitri
and Rall, Patrick
and Yoder, Theodore J.},
title={High-threshold and low-overhead fault-tolerant quantum memory},
journal={Nature},
year={2024},
month={3},
day={01},
volume={627},
number={8005},
pages={778-782},
issn={1476-4687},
doi={10.1038/s41586-024-07107-7},
url={https://doi.org/10.1038/s41586-024-07107-7}
}

@article{Ferris2014tn_qec_decoder,
  title = {Tensor Networks and Quantum Error Correction},
  author = {Ferris, Andrew J. and Poulin, David},
  journal = {Phys. Rev. Lett.},
  volume = {113},
  issue = {3},
  pages = {030501},
  numpages = {5},
  year = {2014},
  month = {7},
  publisher = {American Physical Society},
  doi = {10.1103/PhysRevLett.113.030501},
  url = {https://link.aps.org/doi/10.1103/PhysRevLett.113.030501}
}

@misc{Chubb2021tn_qec_decoder,
      title={General tensor network decoding of 2D Pauli codes}, 
      author={Christopher T. Chubb},
      year={2021},
      eprint={2101.04125},
      archivePrefix={arXiv},
      primaryClass={quant-ph},
      url={https://arxiv.org/abs/2101.04125}, 
}

@Article{Bausch2024ml_qec_decoder,
author={Bausch, Johannes
and Senior, Andrew W.
and Heras, Francisco J. H.
and Edlich, Thomas
and Davies, Alex
and Newman, Michael
and Jones, Cody
and Satzinger, Kevin
and Niu, Murphy Yuezhen
and Blackwell, Sam
and Holland, George
and Kafri, Dvir
and Atalaya, Juan
and Gidney, Craig
and Hassabis, Demis
and Boixo, Sergio
and Neven, Hartmut
and Kohli, Pushmeet},
title={Learning high-accuracy error decoding for quantum processors},
journal={Nature},
year={2024},
month={11},
day={01},
volume={635},
number={8040},
pages={834-840},
issn={1476-4687},
doi={10.1038/s41586-024-08148-8},
url={https://doi.org/10.1038/s41586-024-08148-8}
}

@article{Sweke2021ml_qec_decoder,
doi = {10.1088/2632-2153/abc609},
url = {https://dx.doi.org/10.1088/2632-2153/abc609},
year = {2020},
month = {12},
publisher = {IOP Publishing},
volume = {2},
number = {2},
pages = {025005},
author = {Sweke, Ryan and Kesselring, Markus S and van Nieuwenburg, Evert P L and Eisert, Jens},
title = {Reinforcement learning decoders for fault-tolerant quantum computation},
journal = {Machine Learning: Science and Technology}
}

@article{Battistel2023decoder_review,
doi = {10.1088/2399-1984/aceba6},
url = {https://dx.doi.org/10.1088/2399-1984/aceba6},
year = {2023},
month = {8},
publisher = {IOP Publishing},
volume = {7},
number = {3},
pages = {032003},
author = {Battistel, F and Chamberland, C and Johar, K and Overwater, R W J and Sebastiano, F and Skoric, L and Ueno, Y and Usman, M},
title = {Real-time decoding for fault-tolerant quantum computing: progress, challenges and outlook},
journal = {Nano Futures}
}

@article{Barber2025asic_decoder,
author={Barber, Ben
and Barnes, Kenton M.
and Bialas, Tomasz
and Bu{\u{g}}dayc{\i}, Okan
and Campbell, Earl T.
and Gillespie, Neil I.
and Johar, Kauser
and Rajan, Ram
and Richardson, Adam W.
and Skoric, Luka
and Topal, Canberk
and Turner, Mark L.
and Ziad, Abbas B.},
title={A real-time, scalable, fast and resource-efficient decoder for a quantum computer},
journal={Nature Electronics},
year={2025},
month={1},
day={01},
volume={8},
number={1},
pages={84-91},
issn={2520-1131},
doi={10.1038/s41928-024-01319-5},
url={https://doi.org/10.1038/s41928-024-01319-5}
}

@ARTICLE{Vallero2024understanding,
  author={Vallero, Marzio and Dri, Emanuele and Giusto, Edoardo and Montrucchio, Bartolomeo and Rech, Paolo},
  journal={IEEE Transactions on Quantum Engineering}, 
  title={Understanding Logical-Shift Error Propagation in Quanvolutional Neural Networks}, 
  year={2024},
  volume={5},
  number={},
  pages={1-14},
  doi={10.1109/TQE.2024.3372880}
}

@INPROCEEDINGS{Casciola2022cutting,
  author={Casciola, Nadir and Giusto, Edoardo and Dri, Emanuele and Oliveira, Daniel and Rech, Paolo and Montrucchio, Bartolomeo},
  booktitle={2022 IEEE 28th International Symposium on On-Line Testing and Robust System Design (IOLTS)}, 
  title={Understanding the Impact of Cutting in Quantum Circuits Reliability to Transient Faults}, 
  year={2022},
  volume={},
  number={},
  pages={1-7},
  doi={10.1109/IOLTS56730.2022.9897308}
}

@article{Fowler2015asic,
author = {Fowler, Austin G.},
title = {Minimum weight perfect matching of fault-tolerant topological quantum error correction in average O(1) parallel time},
year = {2015},
issue_date = {January 2015},
publisher = {Rinton Press, Incorporated},
address = {Paramus, NJ},
volume = {15},
number = {1–2},
issn = {1533-7146},
journal = {Quantum Info. Comput.},
month = jan,
pages = {145–158},
numpages = {14}
}

@article{Vallero2025state,
title = {State of practice: Evaluating GPU performance of state vector and tensor network methods},
journal = {Future Generation Computer Systems},
volume = {174},
pages = {107927},
year = {2026},
issn = {0167-739X},
doi = {https://doi.org/10.1016/j.future.2025.107927},
url = {https://www.sciencedirect.com/science/article/pii/S0167739X25002225},
author = {Marzio Vallero and Paolo Rech and Flavio Vella}
}

@misc{leonteva2025benchmark,
      title={Comparative Benchmarking of Utility-Scale Quantum Emulators}, 
      author={Anna Leonteva and Guido Masella and Maxime Outteryck and Asier Piñeiro Orioli and Shannon Whitlock},
      year={2025},
      eprint={2504.14027},
      archivePrefix={arXiv},
      primaryClass={quant-ph},
      url={https://arxiv.org/abs/2504.14027}, 
}

@Article{Ahmadzadeh2024,
author={Ahmadzadeh, Armin
and Sarbazi-Azad, Hamid},
title={Fast scalable and low-power quantum circuit simulation on the cluster of GPUs platforms},
journal={Optical and Quantum Electronics},
year={2024},
month={9},
day={26},
volume={56},
number={10},
pages={1646},
issn={1572-817X},
doi={10.1007/s11082-024-07492-3},
url={https://doi.org/10.1007/s11082-024-07492-3}
}

@article{Gidney2022intrinsicnoisemodel,
   title={Benchmarking the Planar Honeycomb Code},
   volume={6},
   ISSN={2521-327X},
   url={http://dx.doi.org/10.22331/q-2022-09-21-813},
   DOI={10.22331/q-2022-09-21-813},
   journal={Quantum},
   publisher={Verein zur Forderung des Open Access Publizierens in den Quantenwissenschaften},
   author={Gidney, Craig and Newman, Michael and McEwen, Matt},
   year={2022},
   month=sep, pages={813} }


\end{document}